# Numerical Study of Viscoelastic Upstream Instability

**Sai Peng**[1], **Tingting Tang**[1], **Jianhui Li**[1,2], **Mengqi Zhang**[3] **and Peng Yu**[1, 4, 5]†

[1]Department of Mechanics and Aerospace Engineering, Southern University of Science and Technology, Shenzhen, 518055, China

[2]Guangxi Academy of Science, Nanning, 530007, China

[3]Department of Mechanical Engineering, National University of Singapore, 9 Engineering Drive 1, 117575, Singapore

[4]Guangdong Provincial Key Laboratory of Turbulence Research and Applications, Southern University of Science and Technology, Shenzhen 518055, China

[5]Center for Complex Flows and Soft Matter Research, Southern University of Science and Technology, Shenzhen 518055, China

**Abstract**

In this work, we report numerical results on the flow instability and bifurcation of a viscoelastic fluid in the upstream region of a cylinder in a confined narrow channel. Two-dimensional direct numerical simulations based on the FENE-P model (the finite-extensible nonlinear elastic model with the Peterlin closure) are conducted with numerical stabilization techniques. Our results show that the macroscopic viscoelastic constitutive relation can capture the viscoelastic upstream instability reported in previous experiments for low-Reynolds-number flows. The numerical simulations reveal that the non-dimensional recirculation length ($L_D$) is affected by the cylinder blockage ratio ($BR$), the Weissenberg number ($Wi$), the viscosity ratio ($β$), and the maximum polymer extension ($L$). Close to the onset of upstream recirculation, $L_D$ with $Wi$ satisfy Landau-type quartic potential under certain parameter space. The bifurcation may exhibit subcritical behaviour depending on the values of $L^2$ and $β$. The parameter $β$ and $L^2$ have nonlinear influence on the upstream recirculation length. This work contributes to our theoretical understanding of this new instability mechanism in viscoelastic wake flows.

## 1. Introduction

A mixture of a Newtonian fluid and high-molecular-weight polymers (even at extremely low concentration) exhibits viscoelasticity. In the low-Reynolds-number range, when the elasticity becomes the dominant source of nonlinearity, this flow can manifest many interesting and novel flow phenomena that have not been studied extensively, especially on flow instabilities, such as symmetry breaking (Arratia *et al.* 2006; Poole *et al.* 2007; Haward *et al.* 2018, 2020), secondary flow (Yue *et al.* 2008; Davoodi *et al.* 2018), time dependency and even elastic instability and turbulence (Shaqfeh 1996; Larson 2000; Groisman & Steinberg 2000, 2001; Schiamberg *et al.* 2006; Grilli *et al.* 2013; Varshney & Steinberg 2019; Steinberg 2021). These instability phenomena widely exist in processes ranging from plastic manufacturing (Denn 2001; Varchanis *et al.* 2021) to human blood plasma (Brust *et al.* 2013; Thiébaud *et al.* 2014) and porous media flows (Walkama *et al.* 2020; Hopkins *et al.* 2021). However, detailed understandings of the underlying mechanisms in many instances remain vague. In these flows, because the extra elastic stress plays a key role, it is essential to know its value and how it influences the flow field. Experimental measurement and numerical simulation are the two main approaches to obtain the elastic stress. Compared to a large number of reported experimental studies with regards to elastic instability phenomena, numerical method is applied relatively less (Poole *et al.* 2019) to this problem and deserves to be employed more frequently because of its advantages such as more feasible parametric studies and direct analysis of the elastic stress field. However, a notoriously difficult problem in numerical simulation of viscoelastic flow has challenged researchers for decades, namely, the numerical instability when simulating flow at high Weissenberg number (*Wi*), which is called the high Weissenberg number problem (HWNP). The reader is referred to the recent review paper by Alves *et al.* (2021) for a more complete account on these issues. In the following, we will focus on explaining an idealized flow model, that is, viscoelastic flow around a circular cylinder and the recent studies on its upstream instability.

†Email address for correspondence: yup6@sustech.edu.cn

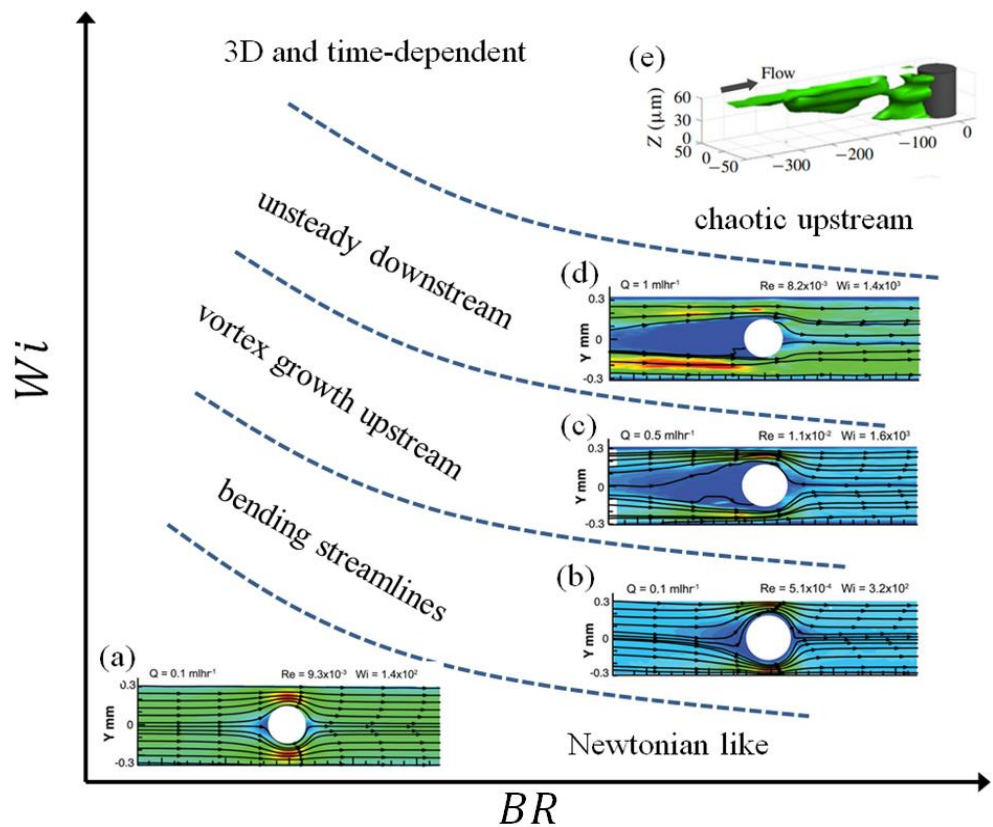

Figure 1. Summary for the *Wi~BR* state space of flow patterns, which is adapted from Zhao *et al.* (2016). Panel (e) is adapted from Qin *et al.* (2019*a*).

Viscoelastic flows past bluff bodies are frequently encountered in industrial applications (e.g., filtration processes and oil extraction) and natural phenomena (e.g., flow in the soil and blood flow in cardiovascular valves and brain tissues) (Larson 1999; Iliff *et al.* 2012; Marsden 2014). In addition, porous media are frequently modelled by ordered and disordered arrays of microfluidic circular cylinders (Walkama *et al.* 2020; Hopkins *et al.* 2021). Placing one or more cylinders in channel or pipe inlet section is often used as a disturbance source to study the channel viscoelastic instability in experiments (Varshney & Steinberg 2018; Qin *et al.* 2019*b*). Viscoelastic flow around a circular cylinder has also been regarded as a benchmark case for numerical and experimental studies (Ultman & Deng 1971; Dhahir &Walters 1989). In Newtonian wake flows, a downstream recirculation zone and vortex shedding can be observed because of the global instability in this flow (Williamson & Roshko 1988; Barkley 2006; Sipp & Lebedev 2007; Tang *et al.* 2020). However, in viscoelastic wake flows, interestingly, an upstream recirculation can form in cylinder wake flows confined between two plane plates.

The recirculation in front of circular cylinder was firstly reported by Kenney *et al.* (2013). Later, Shi *et al.* (2015), Zhao *et al.* (2016), Qin *et al.* (2019*a*), Haward *et al.* (2021) and Hopkins (2022) successively studied this phenomenon experimentally. These experiments were performed using Boger solutions, which are elastic without shear thinning, such as dilute polyethylene oxide (PEO) or Polyacrylamide (PAA) solutions, or using wormlike fluids with a strong shear thinning behaviour, such as cationic surfactant cetyltrimethylammonium bromide (CTAB) and stable hydrotropic salt 3-hydroxy naphthalene-2-carboxylate (SHNC) solutions. Regarding the geometry of the experimental setup, the blockage ratio (*BR*), i.e., the ratio of cylinder diameter to channel width, in these experiments are not less than 50% and the depth-diameter ratio ($\alpha$) is no more than 1.2. Zhao *et al.* (2016) summarized the flow patterns in the *Wi~BR* space with five states: Newtonian like state, bending streamlines, vortex growth upstream, unsteady downstream, and 3D time-dependent chaotic upstream, which are shown in figure 1. Qin *et al.* (2019*a*) found that this flow is inherently three-dimensional and observed symmetry breaking as well as strong upstream propagation effects via elastic waves. At low *BR* (~10%), there also exists a mild upstream instability, with only streamline bending, which was reported by Ribeiro *et al.* (2014), Nolan *et al.* (2016) and Haward *et al.* (2018). Therefore, it seems that high *BR* is one of the conditions for (strong) upstream instability. Besides, the experiments by Shi *et al.* (2016) and Varshney & Steinberg (2017, 2018) also showed that when the cylinders are arranged in tandem at low *BR*, the recirculation zone is likely to appear in the upstream region of the rear cylinders. Few studies quantitatively investigated the length of upstream recirculation. A schematic diagram of the upstream recirculation length (*l*) is shown in figure 2, which is the distance from the foremost end of the upstream recirculation to the foremost end of cylinder. Using the cylinder diameter *D* to normalize *l* as $L_D = l/D$, Zhao *et al.* (2016) obtained a Landau-type behaviour of the dimensionless upstream recirculation length $L_D$ ($L_D$ increasing with *Wi*). In the experiment of Qin *et al.* (2019*a*), $L_D$ is almost linear to the Weissenberg number based on the first normal stress difference $Wi_N = N_1/2\dot{\gamma}\eta(\dot{\gamma})$, where $N_1$ is the first normal stress difference, $\dot{\gamma}$ is the characteristic shear rate defined by the mean centerline velocity, and $\eta(\dot{\gamma})$ is the shear viscosity at this $\dot{\gamma}$. It is worth pointing out that the definitions of the Weissenberg number in these two studies are different.

In the simulations of viscoelastic fluid flow, an additional elastic stress divergence term is often linearly introduced to the right hand side of Navier-Stokes equations. The elastic stress is closed by the constitutive relation with conformation tensor, such as upper-convected Maxwell model (UCM model, Olsson & Yström 1993), Oldroyd-B model (Oldroyd 1950), FENE series models (Herrchen & Öttinger 1997), Phan-Thien and Tanner model (PTT model, Thien & Tanner 1977), etc. The calculated *Wi* is often limited to a low value due to the aforementioned numerical instability. Varchanis *et al.* (2020) simulated the asymmetric viscoelastic flow around a cylinder, which agreed well with their experiment. Their numerical simulation was still limited to a low *BR*.

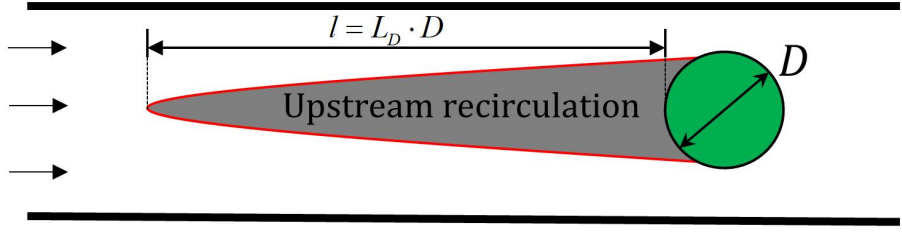

Figure 2. Schematic diagram of upstream recirculation length.

Flow around a cylinder in a channel is a typical mixed flow, including shear flow and tensile flow. Using full-field time-resolved flow induced birefringence (FIB) imaging, Zhao *et al.* (2016) found that upstream instabilities are associated with high stress in the fluid that accelerates in the narrow gap between the cylinder surface and the channel wall when *BR* is large. In the narrowest area between the cylinder surface and the channel wall, the fluid parcels are strongly stretched, which results in high elastic stress there. This sharp growth in elastic stress may cause loss of convergence and trigger the well-known HWNP in numerical simulation (Hulsen *et al.*, 2005). It is a challenging task to balance the numerical accuracy and stability. Up to now, there are no systematic numerical simulations to investigate this flow upstream phenomenon (Poole 2019). In this paper, we present numerical simulations of the viscoelastic upstream instability recently observed experimentally in front of a cylinder in a narrow channel. The finite extensible nonlinear elastic model with the Peterlin closure (FENE-P) is adopted to describe the rheological constitutive behaviour of the dilute polymer solution. The square root reconstruction method (Balci *et al.* 2011) is selected as the stabilization technique for the numerical simulation to handle HWNPs.

The rest of the paper is organized as follows. Section 2 introduces our specific problem, governing equations, solution method, mesh generation, grid convergence test. In Section 3, we discuss our numerical results on the upstream instability. We conclude the present work in Section 4. In the Appendices, we also provide more results on the validation, some additional results on the FENE-CR model, the effect of Peclet number and the effect of different boundary conditions.

2. **Problem formulation and Numerical method**
    **2.1 Problem description**

Figure 3(a) shows the computational domain, which consists of a pressure-driven flow from left to right past a circular cylinder in a confined channel. It is true that Qin *et al.* (2019*a*) revealed the inherent three-dimensional nature of the upstream recirculation zone. However, we will study the corresponding flow instability in this two-dimensional setting as the first attempt. The width of the channel is *H*. The diameter of the cylinder is *D*. Three different blockage ratios (*BR* = *D/H*), i.e., 50%, 62.5% and 75%, are considered. The distance between the cylinder center and the inlet is 37.5*H* (upstream region $l_1$), and the distance between the cylinder center and the exit is 25*H* (downstream region $l_2$). A parabolic flow profile $u(y)=1.5\bar{U}(1-4y^2/H^2)$ is adopted as the inlet velocity, where $\bar{U}$ is the average velocity. The zero-pressure and zero velocity gradient conditions are adopted at the outlet. On the upper wall, the lower wall and the cylindrical wall, the no-slip boundary condition is imposed. The treatment of boundary conditions for the conformation tensor will be discussed below and also in appendix A-4.

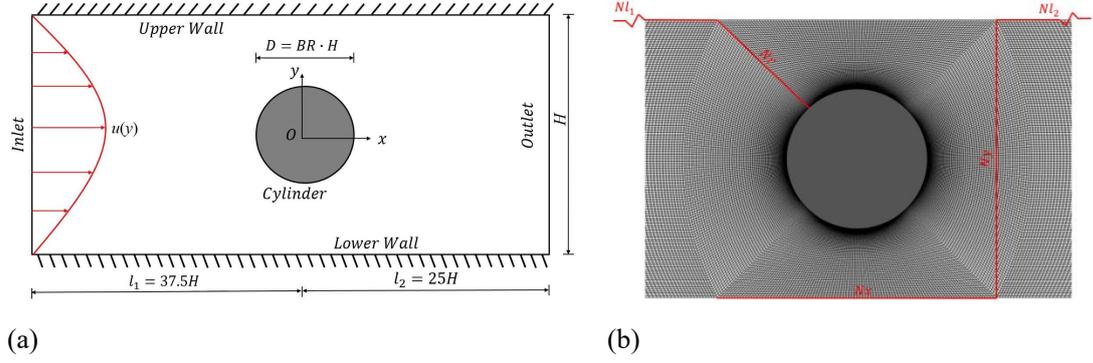

(a)                                              (b)

Figure 3. (a) Schematic diagram of computational domain and (b) the mesh topology near the cylinder.

## 2.2 Governing equations

The viscoelastic fluids in the experiments of Qin et al. (2019) and Pan et al. (2013) show both the shear thickening of elongational viscosity and the shear thinning of the shear viscosity. Thus, the FENE-P model is adopted in the present study, which takes into account the finite elongation of polymer molecules and the bounded stress. The governing equations of the flow combined with the FENE-P constitutive model read as follows (Bird *et al.* 1980; Bird, & Hassager 1987):

$$\nabla \cdot \mathbf{u} = 0,$$
$$\rho \frac{\partial \mathbf{u}}{\partial t} + \rho (\mathbf{u} \cdot \nabla) \mathbf{u} = -\nabla p + \eta_s \Delta \mathbf{u} + \nabla \cdot \boldsymbol{\tau}, \qquad (2.1)$$
$$\frac{\partial \mathbf{c}}{\partial t} + (\mathbf{u} \cdot \nabla) \mathbf{c} - (\nabla \mathbf{u}) \cdot \mathbf{c} - \mathbf{c} \cdot (\nabla \mathbf{u})^T = -\frac{\boldsymbol{\tau}}{\eta_p},$$

with

$$\boldsymbol{\tau} = \frac{\eta_p}{\lambda} \big[ f(\mathbf{c})\mathbf{c} - \theta \mathbf{I} \big], \quad f(\mathbf{c}) = \frac{1}{1 - tr(\mathbf{c})/L^2}, \quad \text{and} \quad \theta = \frac{L^2}{L^2 - 3}. \qquad (2.2)$$

where $\rho$, $\mathbf{u}$, $p$, $t$, $\lambda$, $\mathbf{c}$, $\boldsymbol{\tau}$ and $\mathbf{I}$ are fluid density, flow velocity, pressure, time, the statistical relaxation time of the polymers, conformation tensor, stress tensor of polymer and identity tensor, respectively, $\eta_p$ and $\eta_s$ are the polymeric contribution to zero-shear rate viscosity and the solvent viscosity, respectively, $tr$ denotes trace operator of tensor, the extensibility parameter $L$ measures the maximum stretching of the polymer chains (Bird *et al.* 1980; Bird & Hassager 1987; Purnode & Crochet 1998). When $L$ is set as $\infty$ and $\theta$ tends to 1, the FENE-P model returns to the Oldroyd-B model (Oldroyd 1950).

Note that another common FENE type model is the FENE-CR model, which has been widely applied to predict the flow behaviour of viscoelastic fluid with a constant shear viscosity and a bounded elongational viscosity. We have also performed numerical simulations based on the FENE-CR model. More information about the FENE-CR model and the comparison of the numerical results obtained by the FENE-P model and the FENE-CR model are provided in Appendix A-3.

The main dimensionless parameters for this simulation are the Reynolds number $Re$ and the Weissenberg number $Wi$, which are defined as $Re = \rho \overline{U} H / 2\eta_0$ and $Wi = 2\lambda \overline{U}/H$, respectively. It is emphasized that the present study uses the half channel width as the reference length. Thus, a factor of 2 should be applied to the corresponding $Wi$ defined by using the whole channel width as the reference length when comparing with the published data. $\eta_0 = \eta_p + \eta_s$ is the total viscosity of the solution at zero-shear rate. $\beta = \eta_s/\eta_0$ is the solvent viscosity ratio, a measurement of polymer concentration and molecular characteristics of polymers. In this study, $\beta$ is set to 0.59 in most of the cases. Besides, $\beta = 0.9, 0.75, 0.45, 0.3$ and $0.15$ are also considered in some cases. Four values of $L^2 = 400, 2500, 10000$ and $40000$ are considered. It is suggested from the definitions of $Re$ and $Wi$ that the fluid viscoelasticity is more prominent than the fluid inertia for micro-scale flows. Fluid viscoelasticity can be characterized by $Wi$. In this work, we will mostly only change $Wi$ to probe how the elasticity influences the upstream instability.

The drag coefficient acted on the cylinder is computed as

$$C_d = \frac{1}{\eta_0 \overline{U} H} \int_S [-p\mathbf{I} + \boldsymbol{\tau} + \eta_s \nabla \mathbf{u}] \cdot \mathbf{i} \cdot d\mathbf{S}, \qquad (2.3)$$

where $\mathbf{S}$ is a vector normal to each face element of the cylinder boundary, whose magnitude is equal to the area of face element, and $\mathbf{i}$ is a unit vector aligned with the streamwise direction.

The fluctuation frequency ($f_f$) of flow field and elastic stress wave in the $x$ direction can be obtained by fast Fourier transform (FFT) of the time series of $C_d$. The Strouhal number $St$ is defined as,

$$St = f_f H / (2\overline{U}). \qquad (2.4)$$

In the experiment of Qin *et al.* (2019*a*) and Pan *et al.* (2013), the Weissenberg number is defined using the first normal stress difference. In order to compare our numerical results with their experimental results, we calculate the first normal stress difference of the FENE-P model under simple shear flow as follows (Purnode & Crochet 1998),

$$N_1 = \frac{(1-\beta)\eta_0 L^4}{\lambda(L^2-3)} \left[\frac{2L^6}{\lambda^2 \dot{\gamma}^2 (L^2-3)^2}\right]^{-1/3} \left[\Delta_1^{2/3} + \Delta_2^{2/3} - 2\left(\frac{2L^6}{27\lambda^2 \dot{\gamma}^2 (L^2-3)^2}\right)^{1/3}\right], \qquad (2.5)$$

where,

$$\Delta_1 = 1 + \sqrt{1 + \frac{2L^6}{27\lambda^2 \dot{\gamma}^2 (L^2-3)^2}}, \quad \Delta_2 = 1 - \sqrt{1 + \frac{2L^6}{27\lambda^2 \dot{\gamma}^2 (L^2-3)^2}}.$$

The shear viscosity for the special strain rate ($\dot{\gamma}$) reads

$$\eta = \beta\eta_0 + (1-\beta)\eta_0 \left[\frac{L^6}{4\lambda^2 \dot{\gamma}^2 (L^2-3)^2}\right]^{1/3} \left[\Delta_1^{1/3} + \Delta_2^{1/3}\right]. \qquad (2.6)$$

In experiment, the material properties of a viscoelastic fluid may be complicated and not easy to fit the parameters in the existing viscoelastic constitutive models, including the existence of branched chains of polymer molecules, the spectrum distribution of relaxation time and the certain distribution of chain length of polymer molecules. Herein, we use a normal stress similarity to re-calibrate the rheological parameters. The Weissenberg number based on the first normal stress difference $Wi'_N$ can then be defined as the ratio of the first normal stress difference to twice the Newtonian shear stress (Qin *et al.* 2019a; 2019b), i.e.,

$$Wi'_N = N_1 / \left[ 2\dot{\gamma}(\eta - \eta_s) \right], \qquad (2.7)$$

which is different from $Wi$ for the FENE-P model. The $Wi'_N \sim Wi$ relationship for $\beta = 0.59$ is plotted in figure 4(a). For the Oldroyd-B model, $L$ tends to infinite and we have,

$$Wi'_N = Wi. \qquad (2.8)$$

It is worth mentioning that the above definitions for $Wi$ and $Wi'_N$ do not include the effect of $\beta$. Thus, we adopt another definition proposed in Pan *et al.* (2013) here,

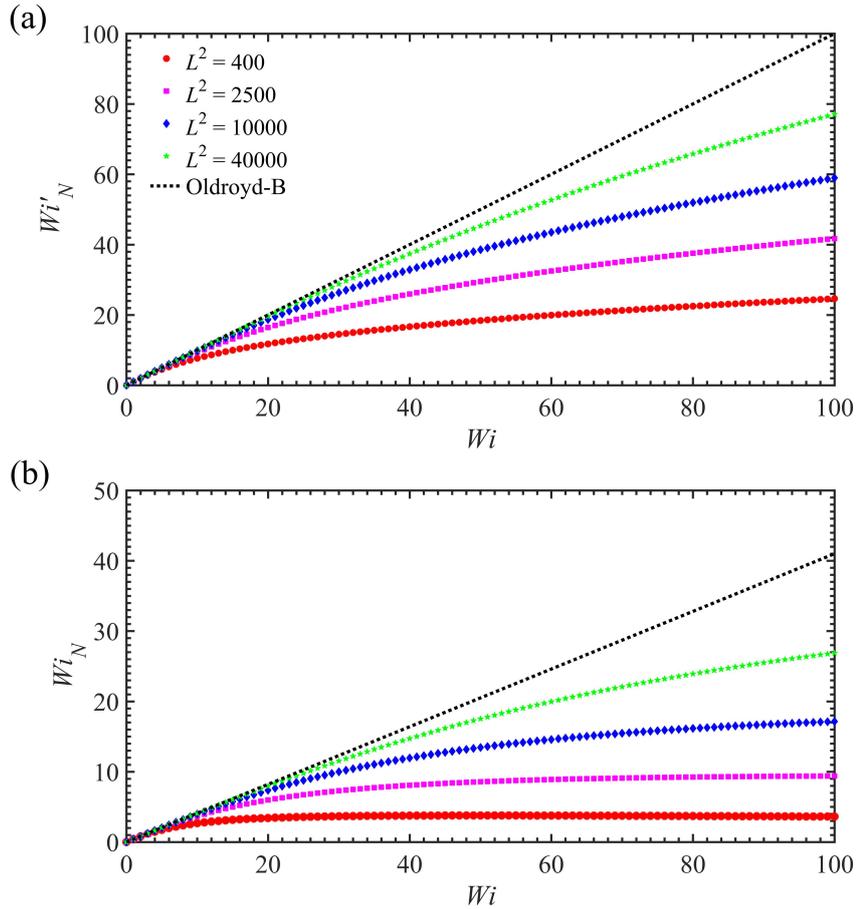

Figure 4. (a) The $Wi'_N \sim Wi$ relationship and (b) the $Wi_N \sim Wi$ relationship for the FENE-P model with $\beta = 0.59$.

$$Wi_N = N_1 / (2\dot{\gamma}\eta). \tag{2.9}$$

Specially, for the Oldroyd-B model with $L \to \infty$, one has,

$$Wi_N = (1-\beta)Wi. \tag{2.10}$$

This definition reflects the influence of $\beta$. $Wi_N$ becomes lower for a larger $\beta$. The corresponding $Wi_N \sim Wi$ relationship for $\beta = 0.59$ is shown in figure 4(b). Note that for $\beta = 0.59$ and $L^2 = 2500$, $Wi_N$ is equal to 9.391 when $Wi = 100$, which is close to the maximum $Wi_N$ in the experiments of Qin et al. (2019a; 2019b) and Pan et al. (2013).

**2.3 Numerical method**

The governing equations are solved by the open-source CFD platform OpenFOAM (Weller et al., 1998) and the rheotool toolbox (Pimenta & Alves 2018). Because no numerical studies have reported the upstream instability in viscoelastic wake flows, we detail our numerical method below and provide extensive validation of the numerical code. In order to ensure the boundedness of $tr(\mathbf{c}) = c_{kk}$ in the FENE-P model, an implicit algorithm is used for pre-calculation before each time step (Richter et al. 2010). Tracing the transport equation of the conformation tensor yields,

$$\frac{\partial c_{kk}}{\partial t} + (\mathbf{u}\cdot\nabla)c_{kk} = tr\left[(\nabla\mathbf{u})\cdot\mathbf{c} + \mathbf{c}\cdot(\nabla\mathbf{u})^T\right] - \frac{(c_{kk}-3)L^4}{\lambda(L^2-c_{kk})(L^2-3)}. \tag{2.11}$$

By defining

$$\varphi = -\ln\left(1 - \frac{c_{kk}}{L^2}\right), \tag{2.12}$$

Eq. (2.11) can be rewritten as,

$$\frac{\partial \varphi}{\partial t} + u_j \frac{\partial \varphi}{\partial x_j} = \frac{e^\varphi}{L^2}\left(c_{kj}\frac{\partial u_k}{\partial x_j} + c_{jk}\frac{\partial u_j}{\partial x_k}\right) + \frac{e^\varphi}{\lambda(L^2-3)}\left(3e^\varphi + L^2 - L^2 e^\varphi\right). \tag{2.13}$$

The scalar $\varphi$ is solved at each time step, and then saved for the calculation of the next time step. Once $\varphi$ is obtained at a given time step, the conformation tensor $c$ can be calculated by Eq. (2.12).

Because of the high singularity of the constitutive governing equations, a small global artificial dissipation term $\kappa\Delta\mathbf{c}$ is added to the right hand side of the transport equation of conformation tensor $\mathbf{c}$ in order to avoid divergence. In our simulations, we set $Re = 0.0001$. Molecular dissipation of polymer should be considered at this $Re$. The presence of this additional diffusive term can be justified by the diffusivity of polymer in solvent, which was estimated to be over the range of $10^{-5}$ to $10^{-7}$ cm$^2$/s (Haggerty et al., 1988). The Schmidt number $Sc = \eta_0/\rho\kappa$, which is defined as the ratio between the zero-shear rate viscosity and the polymer molecular diffusivity, is introduced to quantify the artificial dissipation. For dilute solution of polymer dissolved in water, the density $\rho$ is around 1000kg/m$^3$. The viscosity may vary over a wide range. For example, $\eta_0$ is equal to 0.3 Pa·s in the experiment of Qin et al. (2019a). Thus, $Sc$ is estimated over the range of $10^5$ to $10^8$. The normalized transport equation of conformation tensor is featured by two dimensionless numbers, $Wi$ and the Péclet number ($Pe = Re \cdot Sc$). The value of $Pe$, instead of $Sc$, will be specified in our discussions. In the present study, $Re$ is set to be 0.0001. The corresponding range of $Pe$ is from 10 to 1000. The effect of $Pe$ on the flow behaviour is discussed in Appendix

A-2. By comparing the numerical results and experimental results of Qin et al. (2019a), *Pe* is fixed at 40 for the final simulation.

In order to improve the stability of numerical calculation, stabilization technique must be adopted. More information on common stabilization techniques can be found in Appendix A-1. We perform a preliminary study to test two stabilization techniques, i.e., the logarithmic reconstruction method and the square-root reconstruction method. In the numerical results based on the logarithmic reconstruction method, we do not observe the upstream recirculation found in the experiments of Qin et al. (2019). Moreover, the logarithmic reconstruction method was also adopted by Mokhtari et al. (2022) and Kumar & Ardekani (2022) and the upstream recirculation was not reported in their simulations. However, the square root reconstruction method can successfully predict the upstream recirculation, which is, therefore, adopted in the present simulation. A new symmetric tensor **b** is introduced satisfying $\mathbf{c}=\mathbf{b}\cdot\mathbf{b}^T$. The transport equation of conformation tensor **c** can then be rewritten as,

$$\frac{\partial \mathbf{b}}{\partial t}+(\mathbf{u}\cdot\nabla)\mathbf{b}=\mathbf{b}\nabla\mathbf{u}+\mathbf{ab}+\frac{1}{2\lambda}\left(\theta\cdot(\mathbf{b}^T)^{-1}-\mathbf{b}e^{\varphi}\right)+\frac{\kappa}{2}\Delta\mathbf{b}+\kappa\mathbf{h}, \quad (2.14)$$

where,

$$\mathbf{h}=\mathbf{b}^{-1}\left[\frac{1}{2}(\Delta\mathbf{b})\mathbf{b}+(\partial_x\mathbf{b})^2+(\partial_y\mathbf{b})^2+(\partial_z\mathbf{b})^2\right]. \quad (2.15)$$

As advised by Balci *et al.* (2011), the term $\kappa\mathbf{h}$ in Eq. (2.14) is ignored in our numerical study. Note that **a** in Eq. (2.14) is an antisymmetric tensor, which can be written in the form of components as,

$$\mathbf{a}=\begin{pmatrix} 0 & a_{12} & a_{13} \\ -a_{12} & 0 & a_{23} \\ -a_{13} & -a_{23} & 0 \end{pmatrix}. \quad (2.16)$$

The components of **a** could be calculated by solving the following equations,

$$\begin{aligned} (b_{11}+b_{22})a_{12}+b_{23}a_{13}-b_{31}a_{23}&=w_1, \\ b_{23}a_{12}+(b_{11}+b_{33})a_{13}+b_{12}a_{23}&=w_2, \\ -b_{13}a_{12}+b_{12}a_{13}+(b_{22}+b_{33})a_{23}&=w_3, \end{aligned} \quad (2.17)$$

where,

$$\begin{aligned} w_1&=(b_{12}u_{1,1}-b_{11}u_{2,1})+(b_{22}u_{1,2}-b_{12}u_{2,2})+(b_{23}u_{1,3}-b_{13}u_{2,3}), \\ w_2&=(b_{13}u_{1,1}-b_{11}u_{3,1})+(b_{33}u_{1,3}-b_{13}u_{3,3})+(b_{23}u_{1,2}-b_{12}u_{3,2}), \\ w_3&=(b_{13}u_{2,1}-b_{12}u_{3,1})+(b_{23}u_{2,2}-b_{22}u_{3,2})+(b_{33}u_{2,3}-b_{23}u_{3,3}). \end{aligned} \quad (2.18)$$

where $u_{i,j}$ is the components of $\nabla\mathbf{u}$. For a detailed description of this method, the reader can refer to Balci *et al.* (2011).

The second-order backward scheme is used for time discretization. The time step $\Delta t$ in our simulation is set small enough to ensure numerical stability. The MINMOD scheme is adopted for discretization of $\mathbf{u}\cdot\nabla\mathbf{b}$. The second-order upwind scheme is adopted for discretization of $\mathbf{u}\cdot\nabla\mathbf{u}$. For the tensor **b**, the linear extrapolation boundary with second-order accuracy is imposed on the upper and lower walls, and no-flux boundary condition with first-order accuracy on the cylindrical wall, to ensure the stability of numerical calculation. More discussion on the boundary condition for the tensor **b** can be found in Appendix A-4. The *pUcoupled* algorithm is adopted for the pressure-velocity coupling (Jareteg 2012; Pimenta & Alves 2019). Our simulation is performed using the rheoFoam solver module of rheoTool in OpenFOAM extend 4.0 (Pimenta & Alves 2018). The validation of the present numerical method is provided in Appendix A-1.

The subcritical bifurcation can be distinguished by checking whether the simulation result is path dependence or not, as shown in figure 5. In the present study, we systematically vary the controlling parameter *Wi* and the initial condition to examine their effects on the final simulation result. In the increasing *Wi* process, we gradually and slowly increase *Wi* and use the numerical result of the current state to initialize the simulation for the next simulation. In the decreasing *Wi* process, we gradually and slowly decrease *Wi* and use the numerical result of the current state to initialize the simulation for the next state. We will explicitly point out which process is applied to perform the simulation only if the results at the same *Wi* obtained in the increasing and decreasing *Wi* processes are not identical. For an instability indicator *A* around the linear critical condition (*Wi*$_c$, as shown in figure 5a), finite-amplitude solutions can exist even if *Wi* is less than *Wi*$_c$, which results in a different increasing *Wi* path from a decreasing *Wi* path (the hysteresis phenomenon). This method on identifying subcritical bifurcation is widely used in the previous studies (Becherer *et al.* 2009; Pan *et al.* 2013; Burshtein *et al.* 2017). In this paper, $L_D$, the root mean square upstream recirculation length $RMS(L_D)$, the time-averaged drag coefficient $\overline{C_d}$, the root mean square drag coefficient $C_{drms}$, and an asymmetry parameter *I* (will be defined in Eq. 3.8) are selected as the instability indicators.

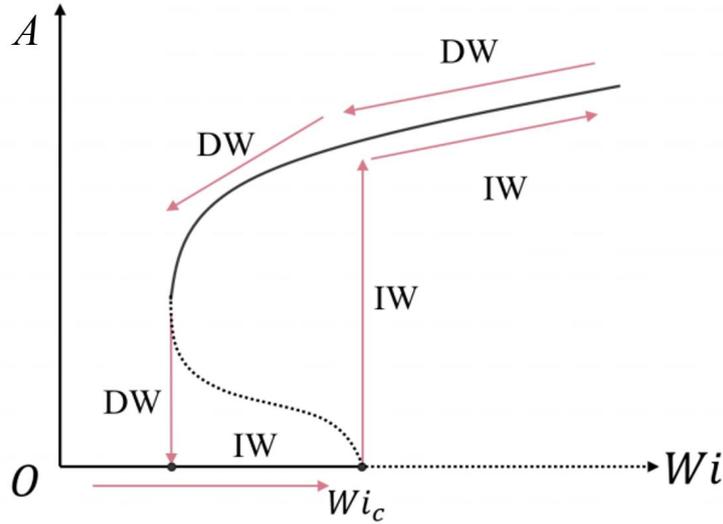

Figure 5. Method to identify subcritical bifurcation. The solid and dashed lines denote the stable and unstable solutions, respectively. *Wi*$_c$ denotes the critical condition. IW and DW denote the increasing and decreasing *Wi* processes, respectively.

**2.4 Mesh generation and grid convergence test**

The block-structured mesh is generated for the computational domain by using the commercial software ANSYS ICEM. The surrounding region of the cylinder is discretized by the O-type mesh, as shown in figure 3(b). The remainder of the computational domain is discretized using several blocks of quadrilateral meshes. In the *y* direction, $N_y$ = 121 (for *BR* = 50%, 201 for *BR* = 65% and 75%) grid points are unevenly distributed. In the $x$ direction, $Nx = Ny$. The dense mesh is applied near the cylinder. The O-type mesh comprises $Ns = 2(Nx + Ny - 2) = 800$ ($BR = 65\%$ and 75%) or 440 ($BR = 50\%$) grid points uniformly distributed along the cylinder perimeter and $Nr = 200$ ($BR = 65\%$ and 75%) or 120 ($BR = 50\%$) grid points stretched over an exponential progression along the radial direction to ensure a fine mesh near the cylinder surface. We set the first cell side-by-side to the cylinder surface in the radial direction to $0.00125H$. In the $x$ direction, $Nl_1 = 701$ grid points are set in the upstream region, and $Nl_2 = 301$ grid points are unevenly arranged in the downstream region. In the $y$ direction, $Ny = 121$ (201 for $BR = 65\%$ and 75%) grid points are unevenly distributed. The total number of cells in the computational domain is approximately 280,000 ($BR = 65\%$ and 75%) or 150,800 ($BR = 50\%$). The detailed mesh generation description can be found in our previous publication (Peng et al. 2021).

The parameter set ($BR$, $Wi$, $L^2$, $\beta$) = (50%, 30, 2500, 0.59) is selected to check the mesh dependency and the corresponding simulation is performed on three sets of meshes, which are listed in table 1. The effect of the mesh size on $L_D$ is investigated. In our numerical simulation, $L_D$ is measured as the horizontal distance from the most remote upstream stagnation point ($u = 0$ and $v = 0$) to the front end of the cylinder. The simulation results show that $L_D$ for Mesh1 is time dependent and its variation range is recorded in table 1. $L_D$ is steady for Mesh2 and Mesh3 and the difference between the results on Mesh2 and Mesh3 is 3%. We thus consider that our numerical results on Mesh2 are accurate enough and the final simulations are performed on this mesh.

| Mesh | $Nl_1$ | $Nl_2$ | $Ny$ | $Nx$ | $Nr$ | $\Delta r$ | Total | $L_D$ |
|---|---|---|---|---|---|---|---|---|
| Mesh1 | 301 | 201 | 51 | 50 | 71 | $0.00125H$ | 39000 | 0.9202~1.0041 |
| Mesh2 | 701 | 301 | 121 | 121 | 71 | $0.00125H$ | 150800 | 1.2329 |
| Mesh3 | 1201 | 501 | 201 | 201 | 121 | $0.000625H$ | 436000 | 1.2726 |

Table 1. ($BR$, $Re$, $Wi$, $L^2$, $\beta$) = (50%, 0.0001, 30, 2500, 0.59). $Nl_1$ and $Nl_2$ denote the grid numbers of the upstream and downstream regions along the $x$ direction, respectively. $Nx$ and $Ny$ denote the grid numbers in the $x$ and $y$ directions, respectively. $Nr$ denotes the grid numbers radiating from the cylinder surface. $\Delta r$ is the grid size of the innermost grid near the cylinder wall.

## 3. Result and Discussions

In this section, we report the numerical results of the viscoelastic wake flow in a confined channel. The parameters considered are ($BR$, $\beta$) = (50%, 0.59) and $L^2$ = 400, 2500, 10000 or 40000 in Section 3.1; $\beta = 0.59$, $L^2$ = 400, 2500, 10000 or 40000 and $BR$ = 10% to 75%, to study the effect of $BR$ in Section 3.2; and $BR$ = 50% or 75%, $L^2$ = 2500 or 10000 and $\beta$ = 0.1- 0.9 to study the effect of $\beta$ in Section 3.3. Table 2 summarizes the parameter space of the present study. In a certain parameter range, the flow is unsteady. We use the data in 5 statistical cycles to calculate the time-averaged PSD and $u_{rms}$ quantities after the flow reaches the statistically steady-state.

Before we perform the flow simulations, the effect of the length of downstream region $l_2$ on the upstream recirculation is examined. Here we consider the parameter set ($BR$, $\beta$, $L^2$, $Wi$) = (50%, 0.59, 2500, 60). Three different $l_2$ of $25H$, $50H$ and $100H$ are investigated. The upstream recirculation lengths at different $l_2$ are summarized in table 3, which indicates that $l_2$ has a very small effect on the upstream recirculation. To save computational time, $l_2 = 25H$ is selected for the final simulation.

| Cases | BR | $\beta$ | $L^2$ | $Wi$ |
| --- | --- | --- | --- | --- |
| Section 3.1 | 50% | 0.59 | 400, 2500, 10000 | 0~100 |
| | 50% | 0.59 | 40000 | 0~60 |
| Section 3.2 | 10%~50% | 0.59 | 2500 | 0~100 |
| | 62.5% | 0.59 | 2500, 10000, 40000 | 0~20 |
| | 75% | 0.59 | 400, 2500, 10000 | 0~10 |
| | 75% | 0.59 | 40000 | 0~9 |
| Section 3.3 | 50% | 0.1-0.9 | 2500 | 0~30 |
| | 75% | 0.3-0.9 | 10000 | 0~20 |
| | 75% | 0.15 | 10000 | 0~15 |

Table 2. Parameter space of the present study.

| $l_2$ | 25$H$ | 50$H$ | 100$H$ |
| --- | --- | --- | --- |
| $L_D$ ($Wi$ =60) | 2.3012 | 2.3001 | 2.2992 |

Table 3. Effect of $l_2$ on the upstream recirculation length.

### 3.1 High-Weissenberg simulations for the FENE-P model at $BR$ = 50%

Simulation is first performed for the case with the parameter set ($BR$, $\beta$, $L^2$) = (50%, 0.59, 2500). ($BR$, $\beta$) = (50%, 0.59) is often used as a benchmark case for numerical simulation. Due to the numerical instability and heavy calculation burden, $Wi$ was often set to less than 0.5 in previous studies (e.g., Fan *et al.* 1999; Alves *et al.* 2001; Hulsen *et al.* 2005). In our numerical simulation, relatively high-$Wi$ flows (up to 100) are calculated by considering polymer molecular diffusion and using a small time step. Lee *et al.* (2021) indicated that the effect of molecular dissipation of polymer can be negligible only when $Pe$ is large ($Pe > $ ~$10^5$) and $Wi$ is small ($Wi < $ ~1). Since the maximum $Wi$ considered is 100 and $Pe$ is equal to 40 in the present study, molecular dissipation of polymer should be taken into account.

The streamline distributions and the $u$-velocity profiles are shown in figure 6. In a Newtonian fluid, the flow is almost symmetry with respect to the cylinder as shown in figure 6(a). As $Wi$ increases, the velocity decreases gradually along the directions approaching the cylinder both upstream and downstream, and the downstream velocity decreases faster. Figure 6(f) indicates that the downstream $u$-velocity almost linearly increases with $x$ except for the region near the cylinder wall for different $Wi$, which is consistent with the experimental results of Haward *et al.* (2018). Note that the strain rate at the rear stagnation point of the cylinder is equal to zero (Haward *et al.* 2019). At $Wi$ = 15, a recirculation zone starts to form upstream of the cylinder. Correspondingly, a region of negative $u$ velocity can be observed in the upstream recirculation region as shown in figure 6(f). With a further increase in $Wi$, the upstream recirculation bubble becomes larger.

To better understand the upstream recirculation, we follow Lee *et al.* (2007) and discuss the flow type parameter defined as

$$\xi(x,y) = \frac{|\dot{\boldsymbol{\gamma}}| - |\boldsymbol{\Omega}|}{|\dot{\boldsymbol{\gamma}}| + |\boldsymbol{\Omega}|}, \tag{3.1}$$

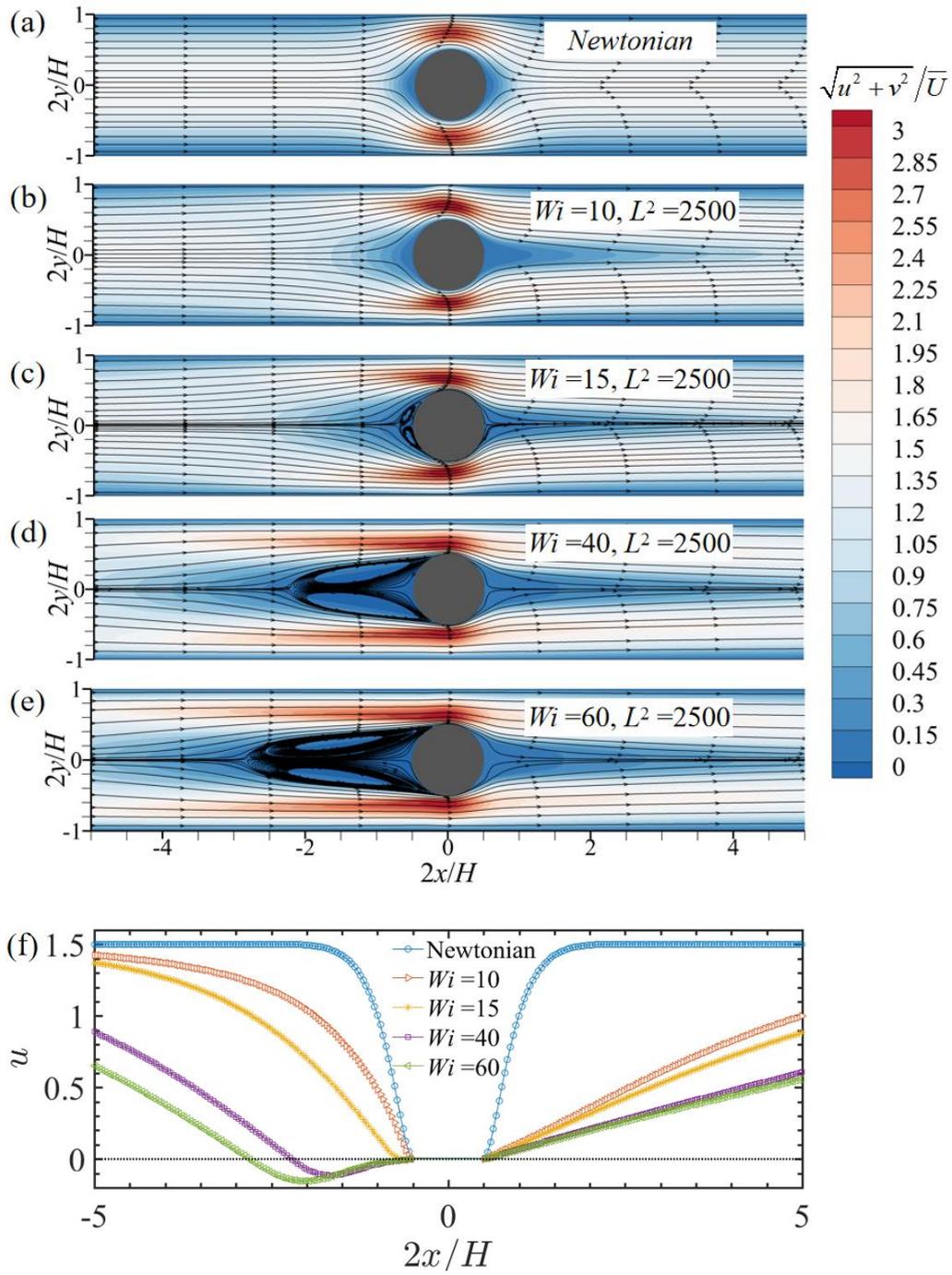

Figure 6. Streamlines for Newtonian (a) and viscoelastic (b-e) fluids at $BR$ = 50%, $\beta$ = 0.59 and $L^2$ = 2500 for different $Wi$ and the corresponding $u$-velocity profiles along $y$ = 0 (f). The contours in (a-e) are colored by $\sqrt{u^2+v^2}/\overline{U}$.

where $|\dot{\boldsymbol{\gamma}}| = \sqrt{\frac{1}{2}\mathbf{D}:\mathbf{D}}$ and $|\boldsymbol{\Omega}| = \sqrt{\frac{1}{2}\boldsymbol{\Omega}:\boldsymbol{\Omega}}$ are the magnitudes of the deformation rate tensor $\mathbf{D} = \frac{1}{2}(\nabla\mathbf{u} + \nabla\mathbf{u}^T)$ and the vorticity tensor $\boldsymbol{\Omega} = \frac{1}{2}(\nabla\mathbf{u} - \nabla\mathbf{u}^T)$, respectively, which can be locally evaluated using the components of the velocity vectors. $\xi$ is all coordinate invariant (Lee *et al.* 2007). Here, $\xi = -1$ indicates solid body rotation, $\xi = 0$ simple shear, and $\xi = 1$ pure extension. The flow strength in the extensional regions is quantified by the principle strain rate or the eigenvector of $\mathbf{D}$ (Astarita 1979),

$$\lambda_1 = \frac{1}{2}\sqrt{(D_{11} - D_{22})^2 + 4D_{12}^2}. \tag{3.2}$$

In two dimensional space, $\lambda_1$ is coordinate invariant. In order to better describe the tensile strength in the extension area, a dimensionless parameter $\lambda_2$ is defined by multiplying $\lambda_1$ and $\xi$,

$$\lambda_2 = \xi\lambda_1. \tag{3.3}$$

Since $\lambda_1$ and $\xi$ are coordinate invariant, $\lambda_2$ is also coordinate invariant. Define a local stretch Weissenberg number as,

$$Wi_{local}^{elastic} = \frac{\lambda}{u^2 + v^2}\left[u^2\frac{\partial u}{\partial x} + uv\frac{\partial v}{\partial x} + uv\frac{\partial u}{\partial y} + v^2\frac{\partial v}{\partial y}\right], \tag{3.4}$$

which describes the local effect on the Weissenberg number caused by stretching (positive $Wi_{local}^{elastic}$) or compression (negative $Wi_{local}^{elastic}$) of fluid parcels along the flow streamline direction.

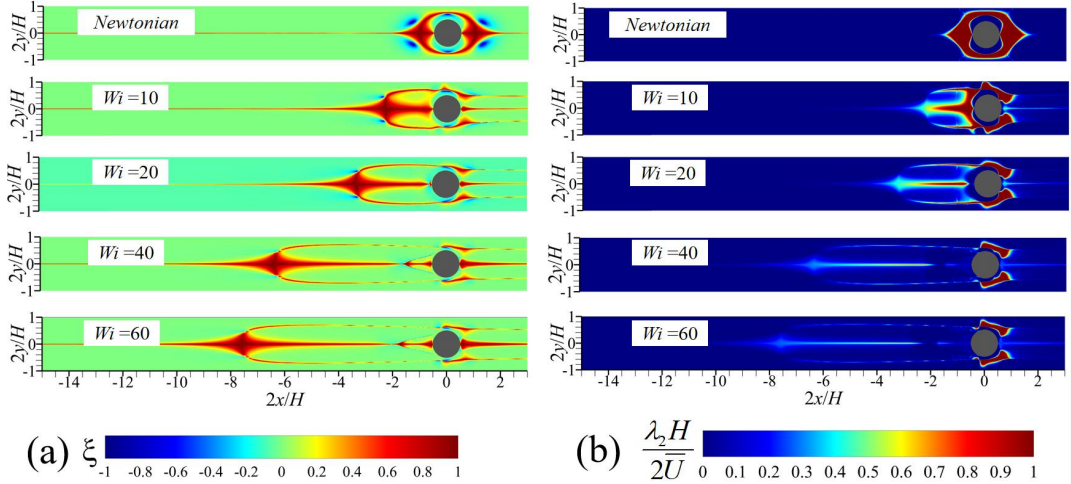

Figure 7. (a) Flow type parameter distributions, and (b) dimensionless $\lambda_2$ distributions.

The $\xi$ and $\lambda_2$ distributions are shown in figure 7(a) and (b), respectively. In a Newtonian fluid, both $\xi$ and $\lambda_2$ distribute symmetrically against the *x* and *y* axes with respect to the cylinder. Fluid parcel around the cylinder exhibits high extension. Just upstream of the cylinder, the flow velocity is reduced, resulting in compression. In the narrow gap between the cylinder surface and the channel wall with $x < 0$, flow velocity increases, resulting in stretching. Figure 7 shows that the upstream compression or stretching regions further expand upstream along the middle line ($y = 0$) and the gap middle line ($y = 3/8H$) as $Wi$ increases. The $\lambda_2$ distribution demonstrates that the strongest tensile strength region appears in the gap between the cylinder surface and the channel wall.

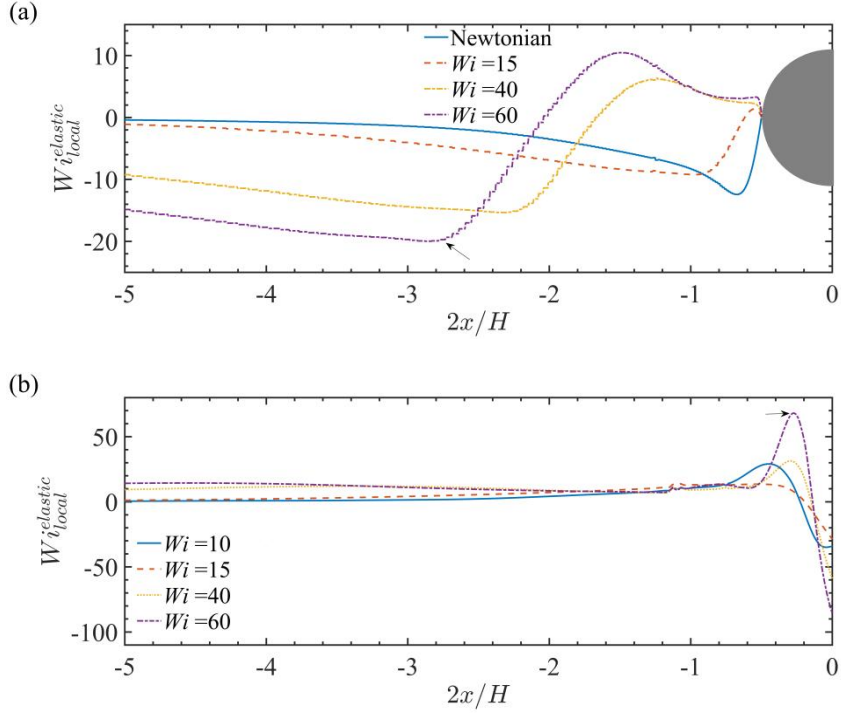

Figure 8. Variation of the local elastic stretch Weissenberg number along the $x$ direction at (a) $y = 0$ and (b) $y = 3/8H$.

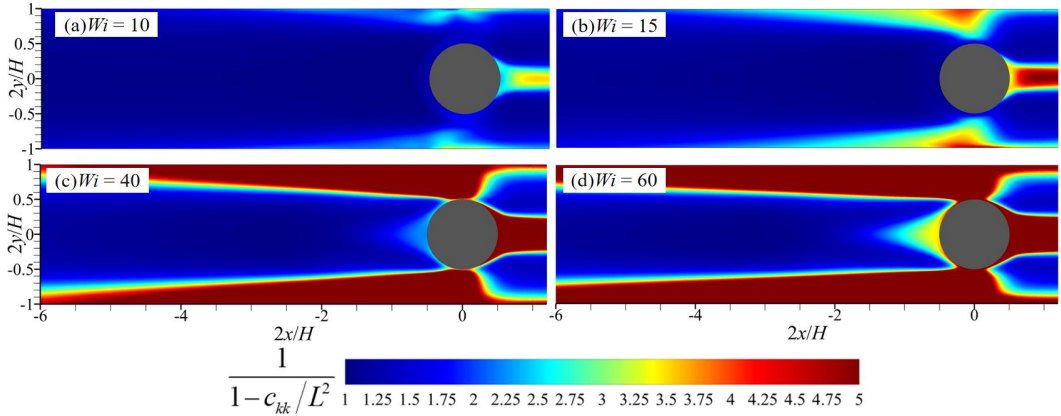

Figure 9. Contours of $1/(1-c_{kk}/L^2)$ for different $Wi$. $1/(1-c_{kk}/L^2)$ is positively correlated with the elastic stress.

The local elastic stretch Weissenberg numbers ($Wi_{local}^{elastic}$) at $y = 0$ and $y = 3/8H$ are extracted and shown in figure 8, respectively. As defined above, the positive and negative $Wi_{local}^{elastic}$ denote fluid stretching and compression, respectively. At $y = 0$, the initial decrease in the flow velocity results in a negative $Wi_{local}^{elastic}$ for a long distance. $Wi_{local}^{elastic}$ initially decreases and then increases along the *x* direction. Generally, the location of the valley point moves upstream and the corresponding minimum $Wi_{local}^{elastic}$ decreases with increasing *Wi*. For *Wi* = 60, the minimum $Wi_{local}^{elastic}$ is about -19.8. At $y = 3/8H$, the increase in the flow velocity results in a positive $Wi_{local}^{elastic}$ for a long distance. Before approaching $2x/H = 0$, *Wi* rapidly increases to the peak value and then experiences a sudden drop. The maximum $Wi_{local}^{elastic}$ is about 68.1 for *Wi* = 60, which occurs in the gap between the cylinder surface and the channel wall. The absolute $Wi_{local}^{elastic}$ in the gap (= 68.1) is much larger than that in front of the cylinder (= 19.8). The large $Wi_{local}^{elastic}$ in the gap reflects that the high elastic stress is concentrated there, which extends upstream near the wall region as shown in figure 9. A similar observation has been reported by Zhao *et al.* (2016) in wormlike fluids.

A well-known dimensionless parameter which rationalizes these streamline instabilities is the *M* parameter introduced by McKinley et al. (1996).

$$M = \left[ \frac{\lambda \overline{U}}{\Re} \frac{\tau_{11}}{(\eta_0 |\dot{\boldsymbol{\gamma}}|)} \right]^{1/2}, \qquad (3.5)$$

where $\lambda \overline{U} = l$ is the characteristic length over which perturbations to the base stress and velocity fields relax, $\Re$ is the streamline radius of curvature, $\tau_{11}$ is the streamwise tensile stress. We will convert this definition of the criterion to a form amenable to local evaluation in flows by the substitution of characteristic values with local values. To consider local effective relaxation time ($\lambda_{eff} = \lambda/f$) and ignore solid-like rotational flows, Cruz *et al.* (2016) proposed a modified Pakdel–McKinley criterion,

$$M^* = \left[ \frac{\lambda_{eff} |\mathbf{u}|}{r} \frac{\tau_{11}}{\|\boldsymbol{\tau}\|_F} \right]^{1/2}, \qquad (3.6)$$

where $r = |\mathbf{u}|^3/|\mathbf{u} \times \dot{\mathbf{u}}|$. Here we note that $\dot{\mathbf{u}}$ is the material derivative of the velocity vector, which is equivalent to $\mathbf{u} \cdot \nabla \mathbf{u}$ for steady-state flow. The Frobenius norm ($\|\cdot\|_F$) is used, so that the resulting $\tau_{11}/\|\boldsymbol{\tau}\|_F$ will vary between zero, when the normal stress is weak, and one, when the tensile normal stress dominates, in highly elastic shear or extensional flows. Therefore in strongly extensional flows, *M\** is approximately given by $\lambda_{eff} |\mathbf{u}|/r$.

Contours of *M\** for different *Wi* are shown in figure 10. The largest *M\** region occurs on the gap between the cylinder and the outer wall. In the upstream recirculation regions, we can also find non-negligible, though smaller, values of *M\**. From the definition of *M\**, high *M\** happens in highly elastic shear or extensional flows. Although the flow curvature in the recirculation zone is large, we only see relatively small value of *M\** in the recirculation region because the extension in this region is not larger than that in the gap regions (figure 8).

The above results indicate that $Wi_{local}^{elastic}$ and the stress dominate in the gap, compared with those in front of the cylinder. In this sense, the flow across the gap can be regarded as a main flow, while the upstream recirculation is a secondary flow. A primary-secondary flow model shown in figure 11 could be applied to characterize this upstream recirculation, i.e., a high-speed stretching gap flow and a relatively low-speed upstream recirculation.

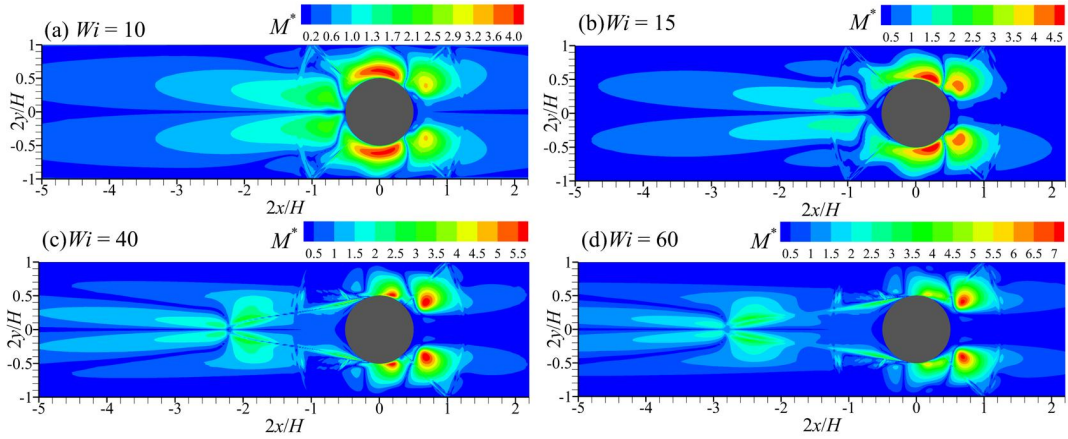

Figure 10. Contours of $M^*$ in the viscoelastic cylindrical wake flow with upstream instability for different $Wi$.

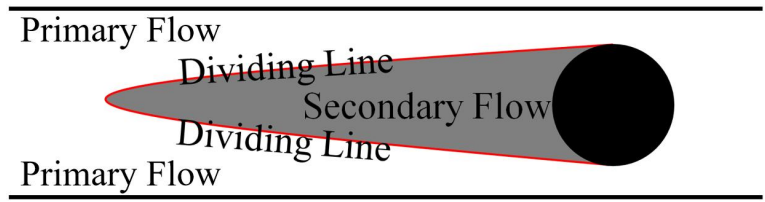

Figure 11. The primary-secondary flow model for the present flow configuration.

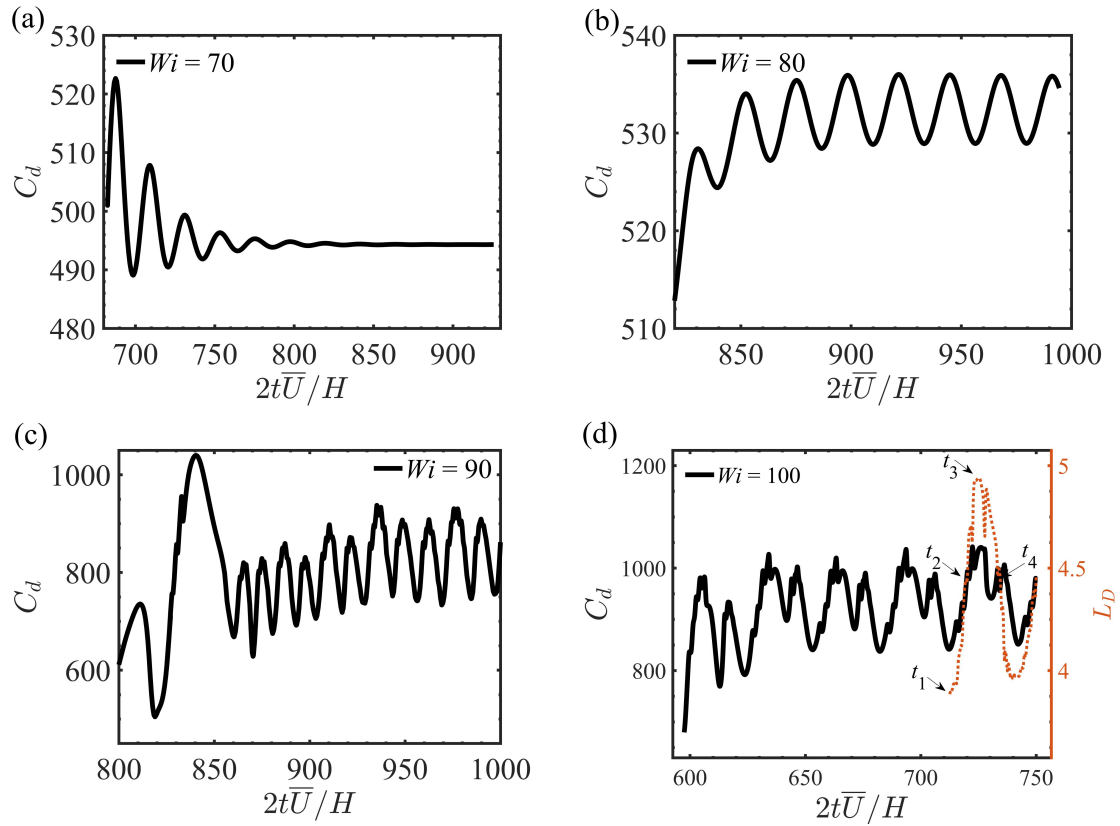

Figure 12. The solid black lines denote the temporal histories of the drag coefficient at $(BR, \beta, L^2)$ = (50%, 0.59, 2500) for (a) $Wi = 70$, (b) $Wi = 80$, (c) $Wi = 90$, and (d) $Wi = 100$. The dash brown line in (d) denotes the temporal history of $L_D$.

The time evolutions of the drag coefficient for $Wi$ = 70, 80, 90 and 100 are plotted in figure 12. To save computing time, the numerical result of $Wi$ = 60 (at time of $2t\overline{U}/H$ = 525) is set as the initial value for these higher-$Wi$ flows. The drag acted on the cylinder increases with $Wi$ when $Wi > \sim 0.5$ (refer to data in table A.1 for low $Wi$ range and data here in high $Wi$ range), no matter whether the flow is steady (Mokhtari et al. 2022) or unsteady. The time-averaged drag coefficient ($\overline{C_d}$) is ~532 at $Wi$ = 80 but ~799 at $Wi$ = 82.5. A sharp increment in $\overline{C_d}$ is observed when $Wi$ increases from 80 to 82.5 as shown in figure 14(a), which results from the enhancement in the additional extensional viscosity due to flow fluctuations (Browne et al. 2021) as shown in figure 14b. A similar sudden increase of drag was also reported in the numerical simulation of Grili et al. (2013). Therefore, the increment in $\overline{C_d}$ with increasing $Wi$ is sharper in the unsteady flow than the steady flow.

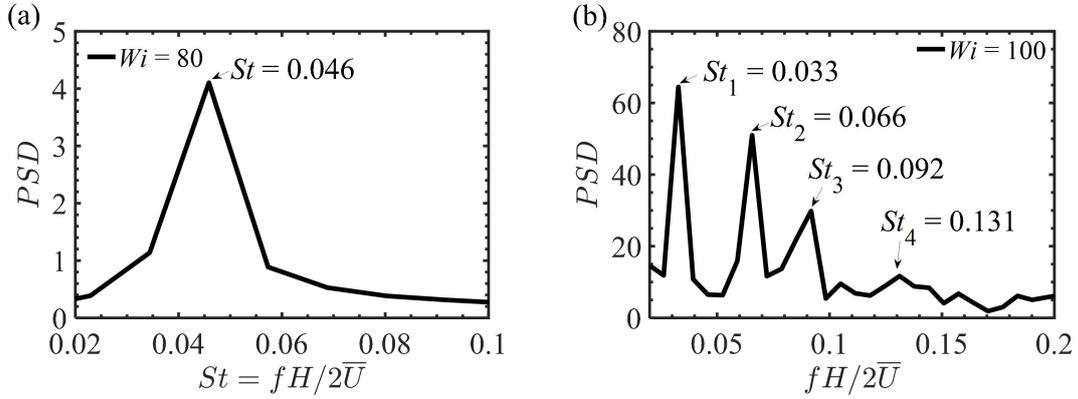

Figure 13. Fast Fourier transform of the drag coefficients for $(BR, \beta, L^2)$ = (50%, 0.59, 2500) at (a) $Wi$ = 80 and (b) $Wi$ =100.

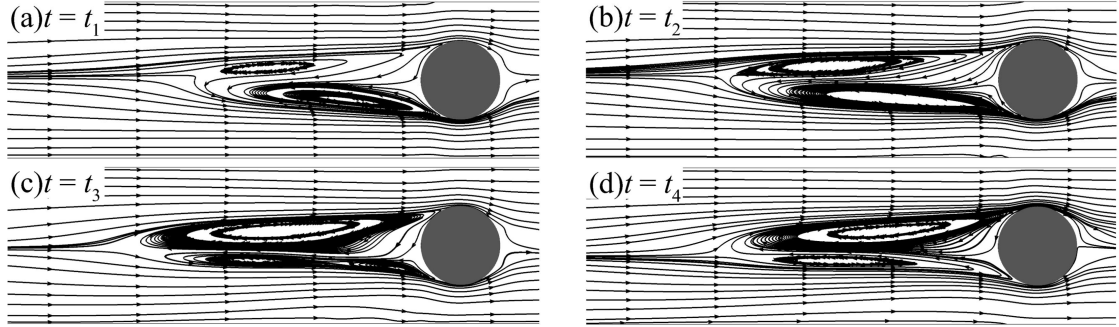

Figure 14. Streamlines at time instants (a) $t = t_1$, (b) $t = t_2$, (c) $t = t_3$, and (d) $t = t_4$ marked in figure 12(d) for $(BR, \beta, L^2, Wi)$ = (50%, 0.59, 2500, 100).

When $Wi < \sim 80$, our calculation indicates that the flow is steady after developing for a long time. However, the flow becomes unsteady when $Wi \geq 80$. At $Wi$ = 80, the fluctuation of drag coefficient exhibits a fully-developed periodic state with a single frequency demonstrated by the FFT analysis shown in figure 13(a). For $Wi$ =100 in figure 12(d), the variation of drag coefficient roughly maintains a periodic state that features more discrete frequencies demonstrated by the FFT analysis shown in figure 13(b). Interestingly, the quantitative analysis demonstrates $St_4 \approx 4St_1$, $St_3 \approx 3St_1$, and $St_2 \approx 2St_1$, indicating the nonlinear effect of the period-doubling. Figure 12(d) also shows the corresponding time evolution of $L_D$ for one period, which exhibits similar trend to the $C_d$ curve. The maximum drag of the cylinder corresponds to the longest upstream recirculation. The corresponding streamlines at four time instants within one period are shown in figure 14.

A sharp increase of drag between from $Wi = 80$ and $82.5$ under the increasing $Wi$ process (i.e., the discontinuity on the $\overline{C_d}$~$Wi$ curve near $Wi = 82.5$) shown in figure 15(a) implies a subcritical transition. We use the result of $Wi = 82.5$ as the initial field to study this transition behavior and gradually reduce $Wi$ (i.e., the decreasing $Wi$ process). It is found that the $\overline{C_d}$~$Wi$ curve maintains continuity until $Wi = 67.5$ and then experiences a sudden drop between $Wi = 67.5$ to $65$ as shown in figure 15(a). Both figures 15 and 16 show that the unsteady flow becomes steady when $Wi$ is decreased from $Wi = 67.5$ to $Wi = 65$. The hysteresis phenomenon shown in figure 15(b) implies that the transient transition is a subcritical Hopf bifurcation.

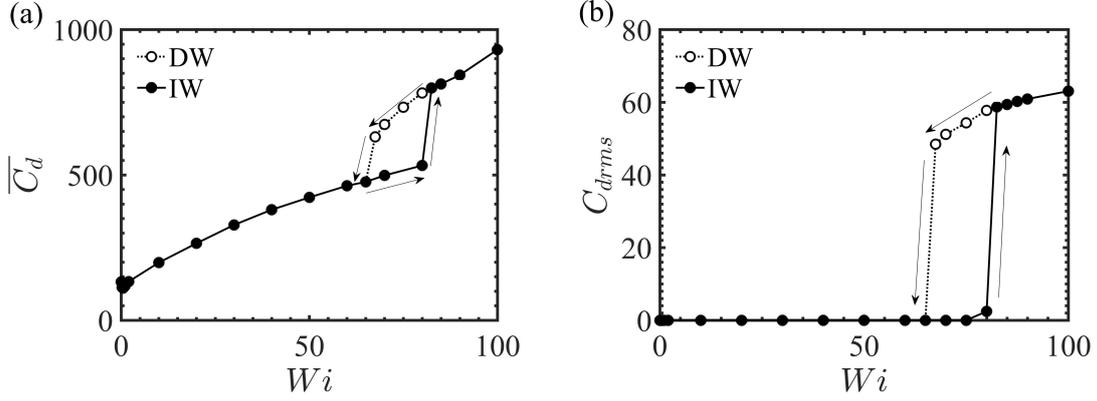

Figure 15. Variations of (a) the time averaged drag coefficient ($\overline{C_d}$) and (b) the root mean square drag coefficient ($C_{drms}$) with $Wi$ for ($BR$, $\beta$, $L^2$) = (50%, 0.59, 2500). IW and DW denote the increasing and decreasing $Wi$ processes, respectively.

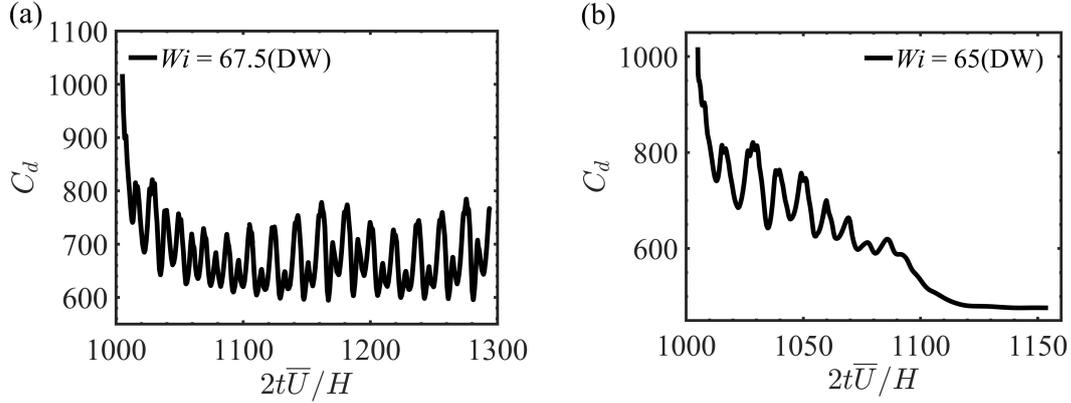

Figure 16. The temporal histories of the drag coefficient at ($BR$, $\beta$, $L^2$) = (50%, 0.59, 2500) for (a) $Wi = 67.5$ and (b) $Wi = 65$ under the decreasing $Wi$ process. The numerical result of $Wi = 82.5$ is selected as the initial field. DW denotes the decreasing $Wi$ process.

For $L^2 = 10000$ and $40000$, we also observe the upstream recirculation. The non-dimensional recirculation length ($L_D$) as a function of $Wi$ for $L^2 = 2500$, $10000$ and $40000$ is summarized in figure 17(a). At high $Wi$, the flow becomes unsteady for $L^2 = 2500$, as discussed above. For $L^2 = 40000$, the flow also becomes unsteady when $Wi \geq 70$. However, for $L^2 = 10000$, the unsteady behaviour is only observed when $20 \leq Wi \leq 50$. When the flow is unsteady and $L_D$ varies with time, the maximum and minimum values of $L_D$ at the corresponding $Wi$ are provided, as shown in figure 17(a). Note that we only present the result up to $Wi = 60$ for $L^2 = 40000$ here. A further increase in $Wi$ causes the simulation divergence due to numerical instability, even we use a very small time step. We observe a sudden shooting up of $L_D$ before the simulation divergence.

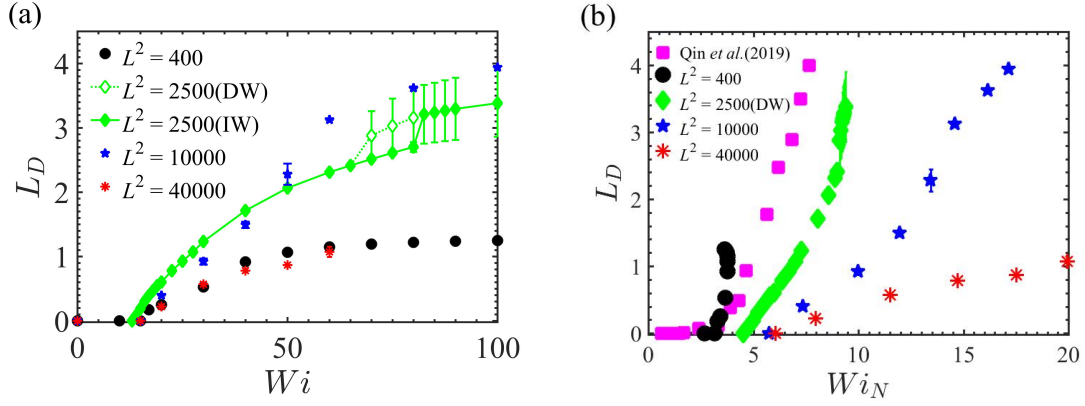

Figure 17. (a) The $L_D \sim Wi$ relationship and (b) the $L_D \sim Wi_N$ relationship at $(BR, \beta) = (50\%, 0.59)$. When the flow becomes unsteady and $L_D$ varies with time at high $Wi$, the time averaged $L_D$ is provided while the error bar indicates the corresponding maximum and minimum values. IW and DW denote the increasing and decreasing $Wi$ processes, respectively.

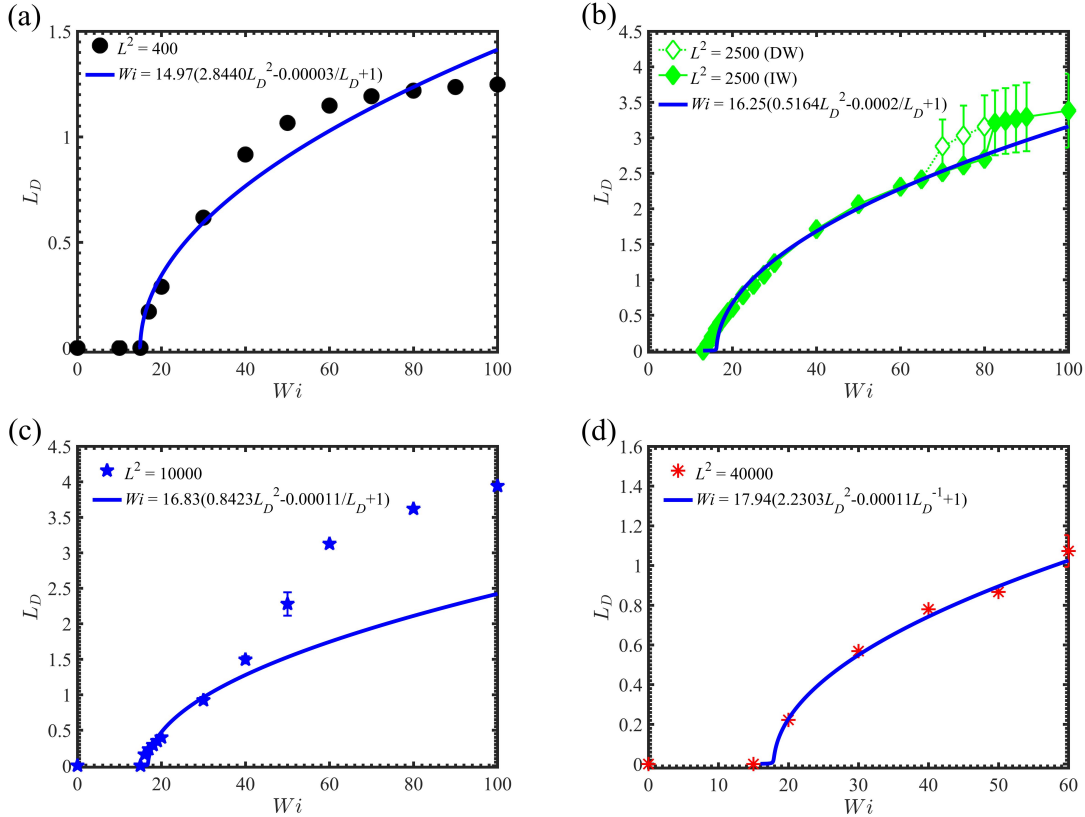

Figure 18. Variation of $L_D$ with $Wi$ at $BR = 50\%$ for (a) $L^2 = 400$, (b) $L^2 = 2500$, (c) $L^2 = 10000$, and (d) $L^2 = 40000$. The solid line denotes the data fitted with Landau-type quartic potential while the symbols ●, ♦, ★, and ∗ denote the numerical results. The fitted $Wi_c$ for each $L^2$ is summarized in table 4. IW and DW denote the increasing and decreasing $Wi$ processes, respectively.

$L_D$ for $L^2 = 40000$ is shorter than those for $L^2 = 2500$ or $L^2 = 10000$. However, Qin *et al.* (2019*a*) implies the upstream recirculation length is positively correlated with extensional viscosity. Variation of $L_D$ with $Wi_N$ is plotted in figure 17(b). As mentioned above, the definition of $Wi_N$ is similar to that in Pan *et al.* (2013). The time averaged $L_D$ measured by Qin *et al.* (2019*a*) is also plotted in figure 17(b). Behaviour of $L_D$ at $L^2 = 2500$ is mostly consistent with the experimental results of Qin *et al.* (2019*a*). It is worth pointing out that in their experiment, the upstream recirculation length varies strongly with time when $Wi_N$ is beyond 4.35. However, the fluctuation of $L_D$ is not very obvious in our simulation when $Wi_N > 13.5$ for $L^2 = 10000$. Note that the present simulation is two-dimensional and the present rheological model and parameters may be different from those for the material used in the experiment. Besides, *Re* may be different between our simulation and their experiment. Thus, the temporal behaviour of $L_D$ deviates from that of Qin *et al.* (2019*a*).

|        | $L^2 = 400$ | $L^2 = 2500$ | $L^2 = 10000$ | $L^2 = 40000$ |
|--------|-------------|--------------|---------------|---------------|
| $Wi_c$ | 14.97       | 16.252       | 16.83         | 17.94         |
| $g$    | 2.8440      | 0.5164       | 0.8423        | 2.2303        |
| $h$    | -0.00003    | -0.00002     | -0.00011      | -0.0001       |

Table 4. The fitting parameters for $BR = 50\%$ in Eq. (3.7)

Figure 17 clearly shows that there is a critical Weissenberg number $Wi_c$ for the onset of the upstream recirculation at each $L^2$. Close to the transition, $L_D$ increases smoothly from zero to a non-zero value with increasing $Wi$, and the relationship between $L_D$ and $Wi$ can be well described by a simple Landau-type quartic potential minimized as:

$$Wi = Wi_c(gL_D^2 + hL_D^{-1} + 1), \qquad (3.7)$$

where $g$ denotes the growth rate coefficient and $h$ quantifies system imperfection that biases a transition to a favored branch. Figure 18 shows the relationship between $L_D$ and $Wi$ for both the numerical results and the fitted data around the onset of the upstream recirculation region. The corresponding values of $g$ and $h$ are listed in table 4. The fitted $Wi_c$ and $g$ increase with $L^2$ while $h$ is always close to zero. When $L^2 = 10000$, the fitted curve only collapses well with the numerical results when $Wi$ is less than 40. This implies that at $BR = 50\%$, variation in $Wi$ may cause inherent change in flow state when $L^2 = 10000$. It may result from the velocity increase in the gap between the cylinder surface and the channel wall. The time-averaged streamlines and contours of the dimensionless $u$ velocity for $Wi = 40$ and 50 are shown in figure 19(a) and (b). The streamlines indicate that the upstream recirculation elongates when $Wi$ increases from 40 to 50. Meanwhile, the contours of the dimensionless $u$ velocity show that the maximum velocity in the gap between the cylinder surface and the channel wall increases from 3.31 at $Wi = 40$ to 3.65 at $Wi = 50$. The maximum time averaged dimensionless $x$-direction velocity in the flow field $\max(\overline{u}/\overline{U})$ is extracted and plotted in figure 19(c). Obviously, a significant increase in velocity exits between $Wi$ = 40 to 50. At high $Wi$, the elastic stress is concentrated near the channel and cylinder walls, which acts to narrow the gap and accelerate the flow velocity there. For example, the maximum velocity in the gap is 3.65 for $(Wi, L^2) = (50, 10000)$, however, is about 3 for a Newtonian fluid. The effect of $BR$ on the upstream recirculation will be discussed in the next section.

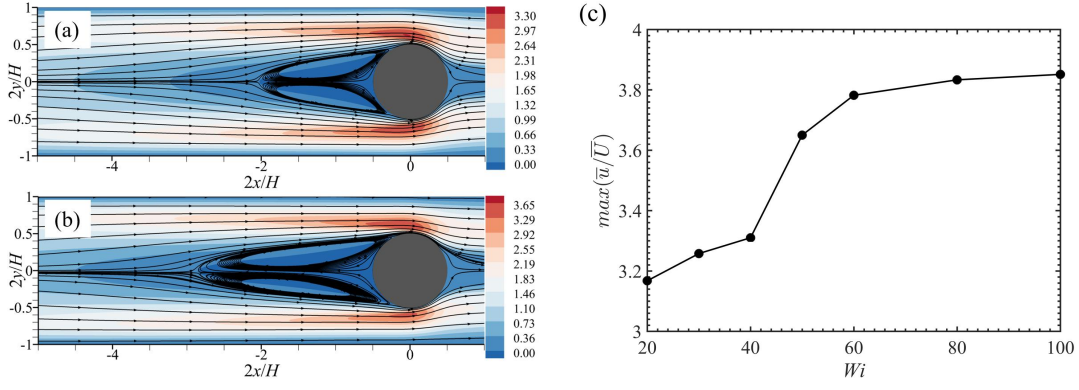

Figure 19 Time-averaged flow streamlines and contours of the dimensionless $u$ velocity at $(BR, \beta, L^2) = (50\%, 0.59, 10000)$ for (a) $Wi = 40$ and (b) $Wi = 50$. (c) Variation of the maximum time-averaged dimensionless $u$ velocity $\max(\overline{u}/\overline{U})$ with $Wi$.

### 3.2. Effect of *BR* on viscoelastic upstream instability

The existing experiments indicated that the upstream recirculation occurs when $BR$ is not less than 50%. When $BR$ is low, such as 10%, only upstream streamline bending exists (Haward *et al.* 2018). Thus, there is a critical value of $BR$ which is around 50% and determined by $Wi$ and $L^2$, below which the upstream recirculation will not occur. When $BR$ is 50% and $Wi$ is low in the simulations discussed in Section 3.1, the velocity in the midline ($y = 0$) changes from 1.5 at the inlet to 0 at the cylinder surface, which results in a compression $Wi_{local}^{elastic}$. However, the velocity along $y = 3/8H$ changes from about 1.5 at the inlet to about 3.0 in the gap, which leads to a stretching $Wi_{local}^{elastic}$. The maximum compression $Wi_{local}^{elastic}$ is almost equal to the maximum stretching $Wi_{local}^{elastic}$ at this situation. When $BR$ exceeds 50%, the gradient of flow velocity in the gap is steeper than that upstream of the cylinder. That is, the stretching $Wi_{local}^{elastic}$ upstream is larger than the compression $Wi_{local}^{elastic}$ in the gap. The compression $Wi_{local}^{elastic}$ in the gap plays a leading role, which results in a primary-secondary flow as shown in figure 11. Moreover, with the increase of $Wi$, the flow velocity inside the gap may slightly increase (discussed in last paragraph in Section 3.1). This assembles that the blockage effect is more severe and the apparent $BR$ is increased (The increase of the flow rate in the center line of the gap can be equivalent to the increase of $BR$). Therefore, the upstream recirculation also occurs when $BR = 50\%$.

It is reasonable to speculate that a certain boundary exists in the space of $(BR, Wi)$ to distinguish the regimes with and without the upstream recirculation. Thus, we studied the effect of $BR$ on $Wi_c$ for $(\beta, L^2) = (0.59, 2500)$, as shown in figure 20. Note that $Wi$ does not exceed 100 in this test. Here $Wi_c$ is not obtained by fitting Eq. (3.7). Instead, at a given $BR$, multiple simulations are performed near $Wi_c$. The bisection method is adopted to determine $Wi_c$ by checking whether the upstream recirculation occurs. Our results imply that the upstream recirculation only occurs when $BR$ is more than 16.7%, i.e., the $Wi_c$ – $BR$ curve asymptotically approaches $BR \sim 16.7\%$.

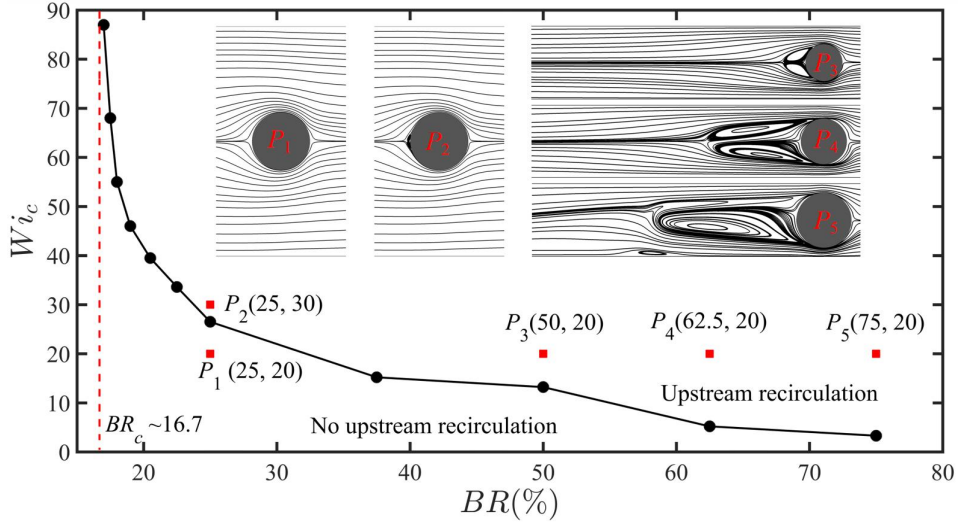

Figure 20. Variation of $Wi_c$ with $BR$ for $(\beta, L^2) = (0.59, 2500)$.

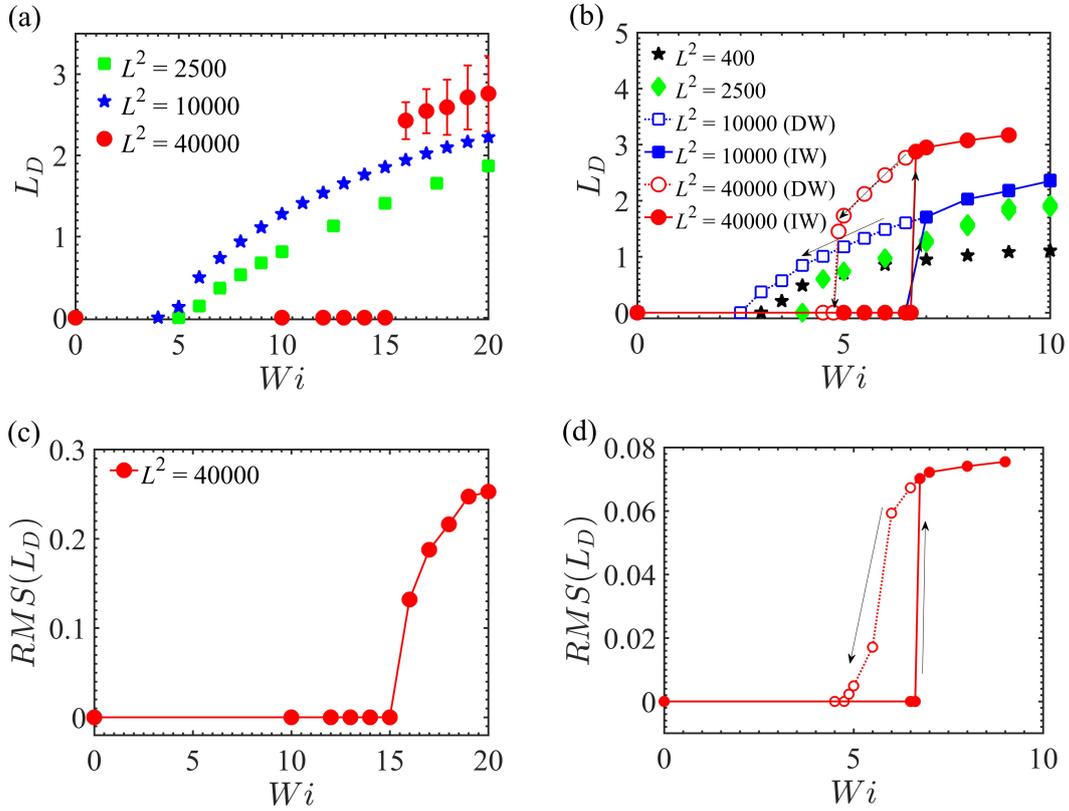

Figure 21. The $L_D \sim Wi$ relationship for (a) $BR = 62.5\%$ and (b) $BR = 75\%$ (for the decreasing Wi process, the numerical results of $Wi = 10$ are selected as the initial fields for $L^2 = 10000$, while the numerical results of $Wi = 9$ are selected as the initial fields for $L^2 = 40000$). Variation in the root mean square $L_D$ with $Wi$ at $L^2 = 40000$ for (c) $BR = 62.5\%$ and (d) $BR = 75\%$. IW and DW denote the increasing and decreasing $Wi$ processes, respectively.

In the rest of this subsection we mainly consider high $BR$ of 62.5% and 75% and discuss its effect on the upstream recirculation. Figure 21(a) and (b) plots $L_D$ as a function of $Wi$ for $BR = $ 62.5% and 75%, respectively. When $(BR, L^2) = (62.5\%, 2500), (62.5\%, 10000), (75\%, 400)$ and (75%, 2500), the growth of upstream recirculation with $Wi$ close to the transition also satisfies Eq. (3.7), as shown in figures 22(a)-(b) and 23(a)-(b). However, when $(BR, L^2)$ = (62.5%, 40000), (75%, 10000), and (75%, 40000), the $L_D \sim Wi$ relationship for the increasing $Wi$ process does not satisfy Eq. (3.7), as shown in figures 21(a)-(b) and 23(d). In fact, there is a sudden shooting up of $L_D$ when $Wi$ is larger than a certain value, indicating that the $L_D \sim Wi$ relationship is discontinuous there. For example, when $BR$ = 62.5% and $L^2$ = 10000, $L_D$ equals zero at $Wi$ = 14.5. However, $L_D$ suddenly changes to 2.186 ~ 2.5697 when $Wi$ > 14.5. Thus, the bifurcation for these parameter sets is subcritical. For decreasing $Wi$, the results for $(BR, L^2)$ = (75%, 10000) and (75%, 40000) still satisfy Eq. (3.7), as shown in figure 23(b) and (c). The flow instability affected by external disturbance. In previous experiments (Kenney *et al.* 2013; Shi *et al.* 2015; Zhao *et al.* 2016; Qin *et al.* 2019a; Haward *et al.* 2021), no sudden shooting-up in the $L_D \sim Wi$ relationship was observed, which may cause by strong disturbance in their experiment. The appearance of upstream recirculation is a more stable form. These scenarios are consistent with our results obtained in the decreasing Wi process, where strong disturbance is also introduces by the initial fields.

Figure 21(c) and (d) plots the root mean square $L_D$ as a function of $Wi$ at $L^2$ = 40000 for $BR$ = 62.5% and 75%, respectively. Comparison between figure 21(a) and 21(c) indicates that, the critical $Wi$ for the onset of the upstream recirculation and the non-zero fluctuation in the upstream recirculation region almost coincide at $Wi$ ~15. The variations of the root mean square $L_D$ with $Wi$ for both the increasing and decreasing $Wi$ processes in figure 21(d) also indicate that bifurcation is subcritical.

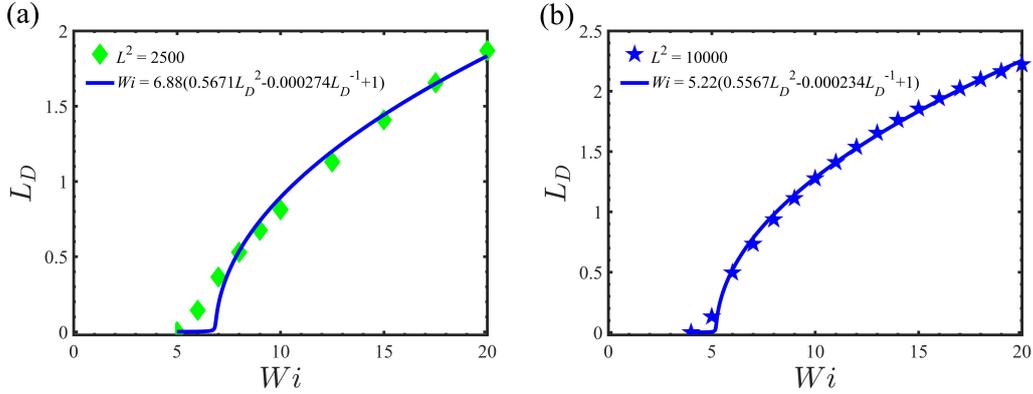

Figure 22. The growth of upstream recirculation with $Wi$ at $BR$ = 62.5% for (a) $L^2$ = 2500, and (b) $L^2$ = 10000. The solid line denotes the data fitted with Landau-type quartic potential while the symbols ♦ and ★ denote the numerical results.

$Wi_c$ for different $BR$ could be obtained from figures 18 and 21. With an increase in $BR$, $Wi_c$ decreases. With an increase in $L^2$, $Wi_c$ becomes larger (the increasing $Wi$ process). Yamani & McKinley (2022) defined an important parameter $Wi/L$ for flow instability. This parameter implies that $L^2$ is positively related to $Wi_c$, which is consistent with the present result. Therefore, the onset of the upstream recirculation is promoted and $Wi_c$ is smaller when $L^2$ is smaller. Moreover, $Wi_c$ is influenced by how the flow field is initialized, i.e., the increasing or decreasing $Wi$ process mentioned in Section 2.3. $Wi_c$ for the decreasing $Wi$ process may be much less than that for the increasing $Wi$ process. For example, $Wi_c$ at $(BR, L^2)$ = (75%, 10000) is about 2.75 for the decreasing $Wi$ process and about 6.75 for the increasing $Wi$ process.

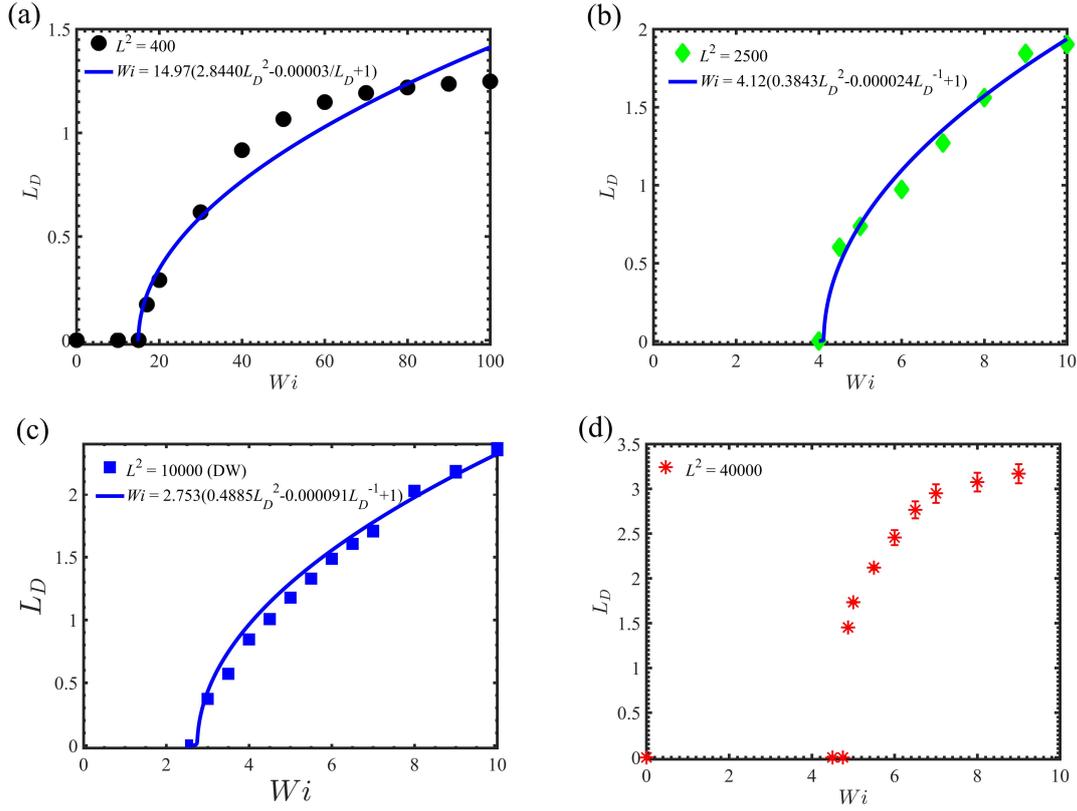

Figure 23. The growth of upstream recirculation with $Wi$ at $BR = 75\%$, for (a) $L^2 = 400$, (b) $L^2 = 2500$, (c) $L^2 = 10000$, and (d) $L^2 = 40000$. The solid line denotes the data fitted with Landau-type quartic potential while the symbols ●, ♦, ■, and ∗ denote the numerical results. DW denotes the decreasing $Wi$ process.

At high $Wi$, such as $(BR, Wi) = (62.5\%, 20)$ or $(75\%, 10)$, $L_D$ only slightly changes with the variation in $Wi$. $L_D$ may reach the saturation value in this situation. For $BR = 62.5\%$ or $75\%$, the saturation value is large when $L^2$ is large, which is different from the trend for $BR = 50\%$. A large $L^2$, i.e., a longer molecular chain length, means higher potential of extensional viscosity. Qin *et al.* (2019*a*) implies the upstream recirculation length is positively correlated with extensional viscosity, which is qualitatively consistent with our results at $BR = 62.5\%$ or $75\%$.

When $BR = 62.5\%$ or $75\%$, the upstream recirculation exhibits strong time dependence behaviour at high $Wi$ when $L^2 = 40000$. The $u$-velocity contours and streamlines at time instant $2t\overline{U}/H = 1710$ for $(BR, L^2, Wi) = (75\%, 40000, 9)$ is shown in figure 24(a). Under this parameter set, the flow is unsteady and the length and shape of the upstream recirculation vary with time. However, the fluctuation amplitude in the upstream recirculation region is not the maximum in the flow field. Instead, the strongest temporal fluctuation occurs in the gap region, as shown by the $u_{rms}$ distribution in figure 24(b). Downstream the cylinder, the high velocity region is concentrated on both sides of the centerline, as shown in figure 24(a). Correspondingly, the flow fluctuation also is high there.

We carefully examine the symmetry of the flow field and find that the symmetry against the horizontal centerline is obviously broken for $BR = 62.5\%$ and $75\%$ when $L^2$ is large, such as $L^2 = 40000$. Figure 25 plots the variations of the averaged velocity (space average) in the upper and lower gaps with time at ($BR$, $L^2$, $Wi$) = (75%, 40000, 9). Both averaged velocity show strongly oscillating behaviour. The corresponding oscillating amplitudes for upper and lower averaged velocities are opposite, in order to satisfy the continuity equation. The upper averaged velocity fluctuates around a mean value of 4.168, which is larger 3.898 for the lower averaged velocity.

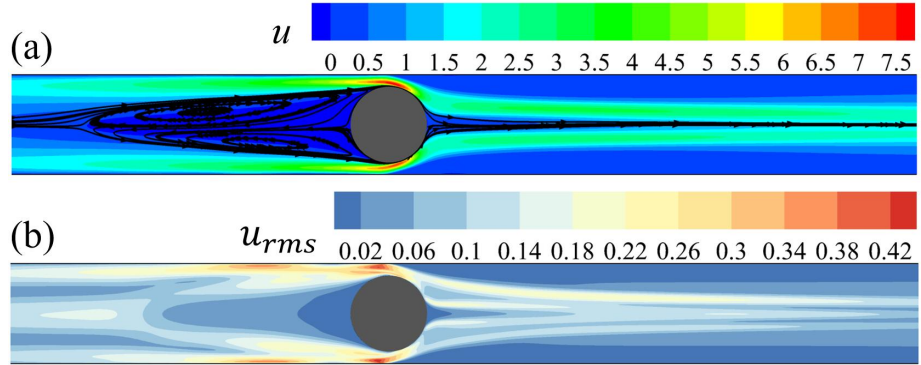

Figure 24. (a) The $u$-velocity contours and streamlines at time instant $2t\bar{U}/H = 1710$, and (b) the corresponding $u_{rms}$ distribution. The parameter set used in this simulation is ($BR$, $L^2$, $Wi$) = (75%, 40000, 9).

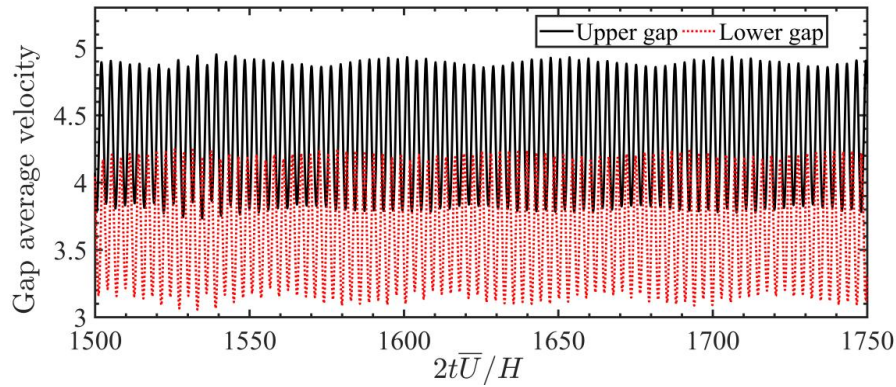

Figure 25. Time histories of the averaged velocity (space average) in the upper and lower gaps for ($BR$, $L^2$, $Wi$) = (75%, 40000, 9).

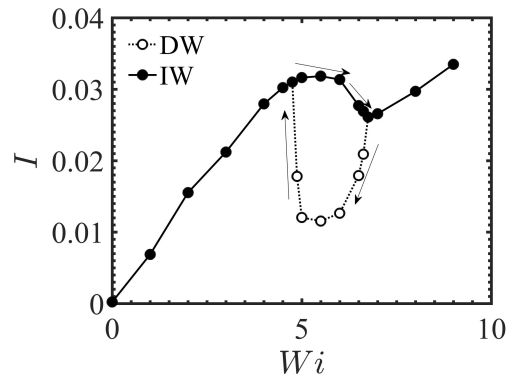

Figure 26. Variation of flow asymmetry parameter $I$ with $Wi$ for ($BR$, $L^2$) = (75%, 40000). IW and DW denote the increasing and decreasing $Wi$ processes, respectively.

A parameter, which follows that of Hopkins et al. (2019), is defined to evaluate the flow asymmetry,

$$I = \frac{|\overline{Q_u} - \overline{Q_l}|}{\overline{Q_u} + \overline{Q_l}}, \tag{3.8}$$

where $\overline{Q_u}$ and $\overline{Q_l}$ are the volumetric flow rate through the upper and lower gaps, respectively. Variation of $I$ with $Wi$ for $(BR, L^2) = (75\%, 40000)$ is presented in figure 26. The flow symmetry only maintains at $Wi = 0$, i.e., $I = 0$ for the Newtonian flow. When $Wi > 0$, the viscoelastic flow exhibits complex asymmetric behavior for high $BR$ and high $L^2$. For the increasing $Wi$ process, $I$ initially increases with $Wi$ until $Wi = 5.5$. Then $I$ decreases with further increase in $Wi$ until $Wi = 6.75$. Note that for $(BR, L^2) = (75\%, 40000)$, the onset of the upstream recirculation occurs at $Wi_c = $ ~6.75. When $Wi > 6.75$, $I$ increases again with $Wi$. For the decreasing $Wi$ process, $I$ decreases with decreasing $Wi$ from $Wi = 9$ to 5.5. However, $I$ increases with a further decrease in $Wi$ until $Wi = 4.75$, which is corresponding to $Wi_c$ below which the upstream recirculation disappears. When $Wi < 4.75$, $I$ decreases again with decreasing $Wi$. The increasing and decreasing $Wi$ paths do not coincide between $4.75 < Wi < 6.75$, which indicates a typical hysteresis phenomenon.

### 3.3. Effect of β and $L^2$ on viscoelastic upstream instability

Figure 18(b) shows that the $L_D$ - $Wi$ relationship for $(BR, L^2) = (50\%, 2500)$ approximately satisfies Landau-type quartic potential near $Wi_c$. However, figure 21(b) the variation of $L_D$ with $Wi$ for $(BR, L^2) = (75\%, 10000)$ under the increasing $Wi$ process does not approximately fit Landau-type quartic potential near $Wi_c$, but shows the hysteresis phenomenon. Thus, these two parameter sets are specially selected to investigate the effect of $\beta$. For $(BR, L^2) = (50\%, 2500)$, $\beta$ ranges from 0.1 to 0.9 and $Wi$ is no more than 30. For $(BR, L^2) = (75\%, 10000)$, $\beta$ ranges from 0.15 to 0.9 and $Wi$ is no more than 20. Within these parameter spaces, the flow fluctuations are small and $L_D$ only varies weakly with time. Thus, the effect of $\beta$ on time-dependent stability is not discussed here.

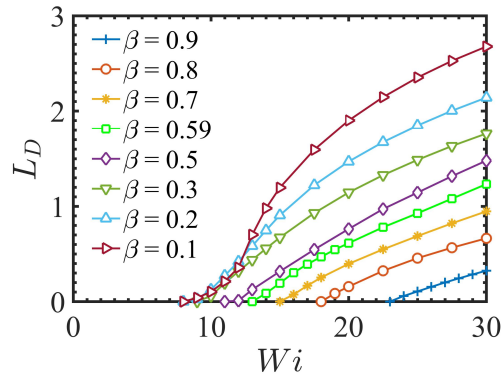

Figure 27. The $L_D \sim Wi$ relationship at $(BR, L^2) = (50\%, 2500)$ for different $\beta$.

In experiments, $\beta$ decreases with increasing polymer concentration. The FENE-P model can describe the shear thinning behaviour of viscoelastic fluid, usually for the case when $L^2$ is small. The smaller $\beta$, the more obvious the shear-thinning effect in the flow. In our study, $L^2$ is large and the influence of $\beta$ on shear-thinning could be ignored (Tamano *et al.* 2020). However, a variation in $\beta$ may affect the distribution of elastic stress as discussed below.

We first consider $(BR, L^2) = (50\%, 2500)$. The variations of $L_D$ with $Wi$ for different $\beta$ are shown in figure 27. Near the upstream recirculation transition point, the $L_D$ - $Wi$ curves all approximately satisfies the Landau-type quartic potential. A smaller $\beta$ corresponds to a smaller $Wi_c$. For example, $Wi_c$ is between 23 to 24 at $\beta = 0.9$, while $Wi_c$ is between 8 to 9 at $\beta = 0.2$. For the same $Wi$, a lower $\beta$ corresponds to a larger $L_D$. For example, $L_D$ is 0.3242 at $\beta = 0.9$ while $L_D$ is 2.1010 at $\beta = 0.2$, for a fixed $Wi = 30$.

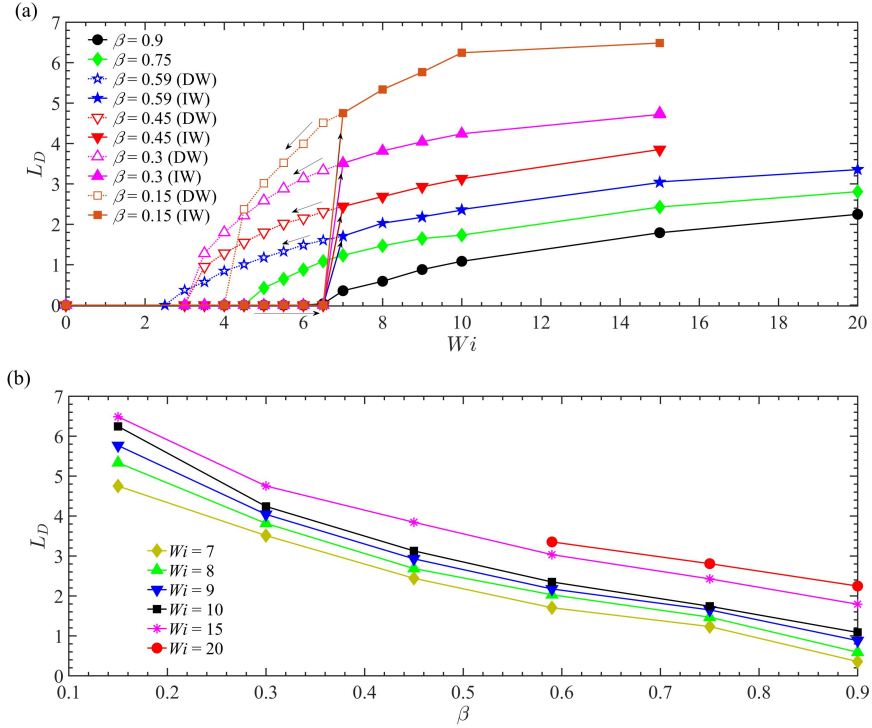

Figure 28. (a) The $L_D$~$Wi$ relationship and (b) the $L_D$~$\beta$ relationship at $(BR, L^2) = (75\%, 10000)$. In panel (b), the larger $L_D$ is selected if $L_D$ obtained in the increasing and decreasing $Wi$ processes (denoted by IW and DW respectively) at the same parameters is not identical.

For $(BR, L^2) = (75\%, 10000)$, $L_D$ for different $(Wi, \beta)$ are counted and shown in figure 28. $\beta$ has a non-linear effect on $L_D$. For both the increasing and decreasing $Wi$ processes, before the onset of upstream recirculation, the $L_D$-$Wi$ relationships are identical and satisfies Eq. (3.7) at $\beta = 0.9$ or 0.75. However, the $L_D$-$Wi$ relationships for the increasing and decreasing $Wi$ processes at $\beta = 0.59, 0.45, 0.3$, or 0.15 are different. For the increasing $Wi$ process, $L_D$ experiences a sudden jump with increasing $Wi$, and the $L_D$-$Wi$ relationship shows a discontinuity. Interestingly, the discontinuity point ($Wi_c$) locates between 6.5 to 7.0 for all $\beta$. These features indicate that the transition is subcritical bifurcation at $\beta = 0.59, 0.45, 0.3$, or 0.15. Thus, we can conclude that $\beta$ affects the bifurcation type, which due to the amplifying effect on elastic stress with decreasing $\beta$. Note that $\beta$ also shows the similar effect in other viscoelastic flows. For example, in the elasto-inertial pipe flow, recent experimental and theoretical studies indicated supercritical bifurcation for large $\beta$ and subcritical bifurcation for small $\beta$ (Choueiri et al. 2021, Chandra et al. 2020, Samanta et al. 2013, Wan et al. 2021). For the decreasing $Wi$ process, all the $L_D$-$Wi$ relationships are still uninterrupted to satisfy Eq. (3.7).

The effect of $L^2$ on the upstream flow behaviour has been discussed scatteredly in the above subsections. We thus provide a brief summary in this subsection. $L^2$ has a noticeable effect of $Wi_c$ for the onset of the upstream recirculation. Generally, a larger $L^2$ corresponds to a higher $Wi_c$ as shown in figures 17(a) and 21(a-b). $L^2$ also affects subcritical behaviour. For large $BR$ with high $L^2$ (such as $L^2 = 40000$ for $BR = 62.5\%$, and $L^2 = 10000$ or $40000$ for $BR = 75\%$), the upstream instability is subcritical. The effect of $L^2$ on subcritical behaviour is similar to that of $1 - \beta$ discussed in section 3.3. For larger $BR$ (e.g. 62.5% and 75%), the unsteady flow is more likely to appear at a larger $L^2$.

## 4. Conclusion Remarks

In this paper, we confirm that the existing macroscopic viscoelastic constitutive relationship (FENE-P model) is still qualitatively applicable to predict the viscoelastic upstream instability through numerical simulation. By applying the square root reconstruction method and considering the molecular dissipation effect to stabilize the numerical simulations, we studied the viscoelastic flow over a circular cylinder in a narrow channel at a very low $Re$ and over a wide range of $Wi$.

The upstream recirculation is observed when $Wi$ is beyond a certain critical value $Wi_c$ and $BR$ is larger than around 16.7%. The occurrence of upstream recirculation is related to high local stretch Weissenberg number $Wi_{local}^{elastic}$ and the high elastic stress in the narrow gap between the cylinder surface and the channel wall, which is consistent with that reported by Zhao *et al.* (2016). For $BR = 50\%$, our simulation results on the upstream recirculation are basically consistent with the experimental results of Qin *et al.* (2019*a*) for the time-averaged behaviour.

Higher $BR$ could precipitate the onset of the upstream recirculation, that is, $Wi_c$ becomes lower. $\beta$ has less effect on $Wi_c$ (the discontinuous point, the increasing $Wi$ process), but has a nonlinearity influence on $L_D$ at high $BR$ (75%) and relative high $L^2$ (10000). For $BR = 50\%$ and $L^2 = 2500$, a lower $\beta$ corresponds to a lower $Wi_c$. Close to the onset of upstream recirculation, $L_D$ with $Wi$ satisfy Landau-type quartic potential when $BR$, $1 - \beta$, and $L^2$ are not very large, e.g., $(BR, \beta, L^2) = (50\%, 0.59, 2500)$. Close to the unsteady transition, the occurrence of the hysteresis phenomenon (signaling a subcritical bifurcation) depends on $L^2$ and $\beta$ (Burshtein *et al.* 2017). For the cases with large $L^2$, small $\beta$ and large $BR$, the flow often exhibits subcritical behavior. In this situation, strong disturbance should be added in the initial flow field so that the corresponding simulation results can be consistent with the experimental results.

When $Wi$ exceeds about a critical $Wi \sim 0.5$ at $BR = 50\%$ (the critical $Wi$ depends on $BR$), the drag on the cylinder increases with $Wi$, although the flow remains steady. When $Wi$ continues to increase beyond a certain threshold, the upstream recirculation becomes unsteady. The drag on the cylinder increases more sharply in this range.

Our results advance the understanding the underlying mechanism of this novel flow instability, which may also be relevant to a number of interesting phenomena observed in viscoelastic flows, e.g., stationary dead zones associated with complex geometries such as multiple tandem circular cylinders (Shi *et al.* 2016) and porous media (Kawale *et al.* 2017).

The cylinder in a channel is often used as a prototype configuration to experimentally study the elastic instability. The upstream instability may extend downstream when $Wi$ is larger. It is interesting to study the elastic wave in the channel through numerical simulations. However, only the elastic instability in the channel without cylinder was studied (Morozov 2022). Our simulation could provide a platform for flows in the channel with cylinder(s).

The upstream recirculation phenomena are governed by many parameters and the present study does not explore all the parameter space of ($BR$, $Re$, $Wi$, $L^2$, $\beta$). These parameters need to be investigated in more detail. In experiment of Shi *et al*. (2015), the upstream instability eventually occurs when the elastic Mach number $Ma = \sqrt{Re \cdot Wi}$ is beyond ~10 at $Re \sim 1$, which is affected by both channel geometry and fluid properties. We believe that $Re$ is an important dimensionless number for such instability. However, the effect of $Re$ is not discussed in this paper and will be studied in our future work.

**Declaration of Interests**

The authors report no conflict of interest.

**Acknowledgement**

The author P Yu would like to thank the financial support from Shenzhen Science and Technology Innovation Commission (Grant No. JCYJ20180504165704491), Guangdong Provincial Key Laboratory of Turbulence Research and Applications (Grant No. 2019B21203001), the National Natural Science Foundation of China (NSFC, Grant No. 12172163, 12002148, 12071367). This work is supported by Center for Computational Science and Engineering of Southern University of Science and Technology. The authors would like to thank Prof. Victor Steinberg from Weizmann Institute of Science and Jia-yu Li from Southern University of Science and Technology for the useful discussions.

**Appendix**

### A-1. Validation

To validate the present numerical method, a series of simulations based on the Oldroyd-B model are performed for the flow past cylinder in a channel with $BR$ = 50% (Mesh2) and the numerical results are compared with published data. In these simulations, the Reynolds number is set as zero, i.e., the convection terms in the momentum equations are ignored. Note that the polymer molecular dissipation is also not considered in this section. The viscosity ratio $\beta$ is fixed at 0.59 while $Wi$ is varied from 0 to 0.55. At low $Wi$, the viscoelastic equations could be numerically solved directly. However, previous numerical tests (Fan *et al*. 1999; Hulsen *et al*. 2005; Alves *et al*. 2001, 2009) indicates that the stabilization technique must be adopted to avoid simulation divergence at high $Wi$. Commonly used stabilization techniques include the elastic viscous split stress (EVSS) method (Fan *et al*. 1999), the log-conformation representation method (Labelled as Log, Fattal and Kupferman 2004; Hulsen *et al*. 2005; Afonso *et al*. 2009) and the square root reconstruction method (Labelled as Sqrt, Alves *et al*. 2009). The present study adopts the square root reconstruction method.

The values of $C_d$ for different $Wi$ are compared with the published data and listed in table A1. Our results are in good agreement with those reported in the literature. For example, the drag coefficient for $Wi$ = 0.35 obtained in our simulation is $C_d$ = 117.33, which is consistent with $C_d$ = 117.315 in Hulsen *et al*. (2005) and $C_d$ = 117.32 in Fan *et al*. (1999). Besides, the stress profiles along the upper cylinder wall and the downstream center line for $Wi$ = 0.3 and 0.45 are shown in figure A1. We can see a good agreement between our result and that in Aleves *et al.* (2001) for *Wi* = 0.3. However, a difference among our results and those of Alves *et al.* (2001) and Afonso *et al.* (2009) can be observed in the region immediately behind the cylinder for *Wi* = 0.45, even the local refined mesh is utilized there in our simulation as shown in figure A1(b). Note that a higher peak of the local elastic stress occurs behind the cylinder when *Wi* is larger, which is very difficult to be accurately resolved by numerical simulation. Anyway, the comparison indicates that our simulation could capture the main flow characteristics downstream since the behaviour of the local elastic stress downstream is similar for the three simulations.

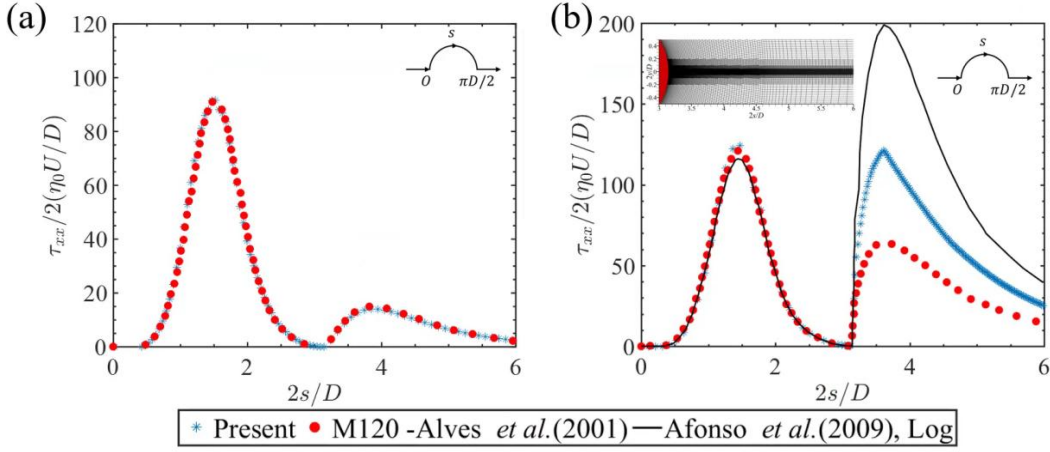

Figure A1. Stress profile along the upper cylinder wall and the downstream center line for the Oldroyd-B fluid at (a) *Wi* = 0.3 and (b) *Wi* = 0.45.

| Wi | Hulsen *et al.* (2005) Log | Fan *et al.* (1999) EVSS | Alves *et al.* (2001) | Afonso *et al.* (2009) Log | Afonso *et al.* (2009) Sqrt | Present Sqrt |
|---|---|---|---|---|---|---|
| 0 | 132.358 | 132.36 | 132.378 | - | - | 132.36 |
| 0.05 | 130.363 | 130.36 | 130.355 | - | - | 130.30 |
| 0.1 | 126.626 | 126.62 | 126.632 | - | - | 126.565 |
| 0.15 | 123.193 | 123.19 | 123.210 | - | - | 123.133 |
| 0.2 | 120.596 | 120.59 | 120.607 | - | - | 120.555 |
| 0.25 | 118.836 | 118.83 | 118.838 | 118.818 | 118.821 | 118.808 |
| 0.3 | 117.775 | 117.78 | 117.787 | 117.774 | 117.776 | 117.776 |
| 0.35 | 117.315 | 117.32 | 117.323 | 117.323 | 117.324 | 117.33 |
| 0.4 | 117.373 | 117.36 | 117.357 | 117.364 | 117.370 | 117.36 |
| 0.45 | 117.787 | 117.80 | 117.851 | 117.817 | - | 117.776 |
| 0.5 | 118.501 | 118.49 | 118.518 | 118.680 | - | 118.508 |
| 0.55 | 119.466 | - | - | 119.780 | - | 119.51 |

Table A1. Variation of $C_d$ with *Wi* at *Re* = 0 for the Oldroyd-B fluid.

### A-2. Effect of *Pe*

In this section, the effect of *Pe* on the upstream flow behavior is examined. The parameter set investigated is (*BR*, *β*, $L^2$, *Wi*) = (50%, 0.59, 2500, 50). Five *Pe* values of 10, 40, 64, 80 and 400 are considered. Figure A2 shows that, when *Pe* is equal to 10, 40 or 64, the flow field is similar to the experimental results of Qin et al. (2019). However, when $Pe$ is set to be 80 or 400, the upstream recirculation becomes short and one or two recirculation regions attached to the channel wall appear. Thus, we select a relative lower $Pe$ = 40 in the final simulation.

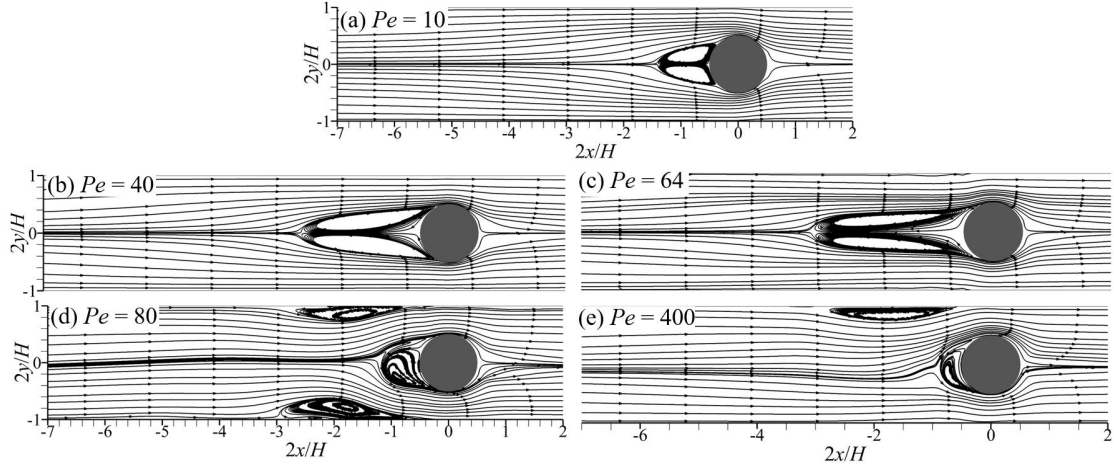

Figure A2. Instantaneous streamlines for $(BR, \beta, L^2, Wi)$ = (50%, 0.59, 2500, 50) at (a) $Pe$ = 10, (b) $Pe$ = 40, (c) $Pe$ = 64, (d) $Pe$ = 80 and (e) $Pe$ = 400.

### A-3. The FENE-CR model

The governing equations based on the FENE-CR model are similar to those for the FENE-P model (Eqs. 2.1 to 2.3) except for the polymer stress definition, which is expressed as,

$$\boldsymbol{\tau} = \frac{\eta_p}{\lambda} f(\mathbf{c})(\mathbf{c} - \mathbf{I}). \tag{A.1}$$

The FENE-CR model is degenerated to the Oldroyd-B model if $f(c)$ is set to 1 in Eq. (A.1).

Tracing the transport equation of the conformation tensor $c$ for the FENE-CR model yields,

$$\frac{\partial c_{kk}}{\partial t} + (\mathbf{u} \cdot \nabla) c_{kk} = tr\left[(\nabla \mathbf{u}) \cdot \mathbf{c} + \mathbf{c} \cdot (\nabla \mathbf{u})^T\right] - \frac{(c_{kk} - 3)L^2}{\lambda(L^2 - c_{kk})}. \tag{A.2}$$

By using the scalar $\varphi$ defined in Eq. (2.12), Eq. (A.2) can be rewritten as,

$$\frac{\partial \varphi}{\partial t} + u_j \frac{\partial \varphi}{\partial x_j} = \frac{e^\varphi}{L^2}\left(c_{kj}\frac{\partial u_k}{\partial x_j} + c_{jk}\frac{\partial u_j}{\partial x_k}\right) + \frac{e^\varphi}{\lambda L^2}\left(3e^\varphi + L^2 - L^2 e^\varphi\right). \tag{A.3}$$

When the square root reconstruction method is applied, the transport equation of conformation tensor $\mathbf{c}$ for the FENE-CR model can then be rewritten as,

$$\frac{\partial \mathbf{b}}{\partial t} + (\mathbf{u} \cdot \nabla)\mathbf{b} = \mathbf{b}\nabla\mathbf{u} + \mathbf{ab} + \frac{1}{2\lambda}\left((\mathbf{b}^T)^{-1} - \mathbf{b}\right)e^\varphi + \frac{\kappa}{2}\Delta\mathbf{b} + \kappa\mathbf{h}. \tag{A.4}$$

All other related equations are the same as those for the FENE-P model and thus not repeated here. The simulation based on the FENE-CR model is also performed using the rheoFoam solver module of rheoTool in OpenFOAM extend 4.0 (Pimenta & Alves 2018).

The parameter set considered here is $(BR, \beta, L^2)$ = (50%, 0.59, 2500). Haward et al. (2019) suggested that the purely elastic instability is only observed when both shear-thinning and elasticity exist. The simulation based on the FENE-CR model would help us to identify whether shear-thinning is necessary for the upstream instability at high *BR*. Variation of $L_D$ with *Wi* for the results based on both the FENE-CR and FENE-P models is shown in figure A3. The results indicate that the simulations based on both the models can predict the occurrence of the upstream recirculation. Although deviations can be observed for the $L_D$ - *Wi* curves based on the two models, the general trends are similar. Therefore, shear-thing is not a necessary condition for this upstream instability.

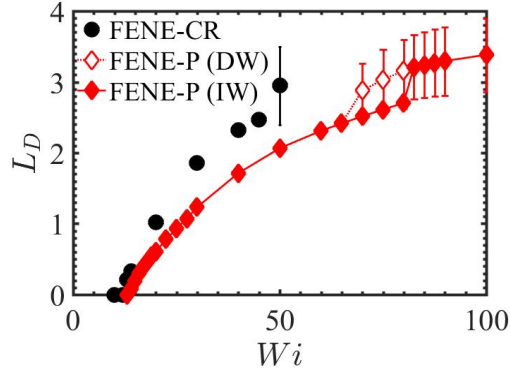

Figure A3. Variation of $L_D$ with *Wi* for (*BR*, *β*, $L^2$) = (50%, 0.59, 2500). The results based on both the FENE-CR and FENE-P models are provided. IW and DW denote the increasing and decreasing *Wi* processes, respectively.

### A-4. Boundary condition for the tensor b

When simulating the viscoelastic fluid flow in OpenFOAM, the boundary condition for the elastic stress on the wall needs to be explicitly implemented, in order to obtain the stress gradient of the center point of a grid element close to the wall (Alves et al, 2021). In our numerical simulation, the **b** tensor adopts the linear extrapolation boundary condition on the channel walls and the no flux boundary on the cylinder wall. These uneven implementations are based on the following considerations. First, if the linear extrapolation boundary condition is imposed on the cylinder wall, a very small time step should be used in the simulation. For example, for the parameter set (*BR*, *β*, $L^2$) = (75%, 0.59, 40000), the dimensionless time step must be set as 6.25 × $10^{-5}$. It takes more than one month for the simulation to reach statistically steady state with the parallel computing on 40 CPU-cores. Second, if the linear extrapolation boundary condition is imposed on the cylinder wall, the simulation results show unstable trend. For example, the upstream recirculation behaves symmetrically in the experiment of Haward et al. (2021). However, the corresponding flow field at (*BR*, *β*, $L^2$, *Wi*) = (75%, 0.59, 40000, 9) exhibits obvious asymmetric behaviour, which implies that too strong disturbance is introduced into the simulation. Finally, if the no-flux boundary condition is imposed on both the channel walls and the cylinder wall, the upstream recirculation only occurs at very high *Wi* in the simulation. The deviation between the numerical and experimental results is unacceptable.

Figure A4 shows the variations of $L_D$ and the root mean square $L_D$ with *Wi* at (*BR*, *β*, $L^2$) = (75%, 0.59, 40000) by using different boundary conditions for the tensor **b** on the cylinder wall. The boundary condition has a noticeable effect on $Wi_c$ and the overall upstream recirculation behaviour. The time averaged *u*-velocity contours and streamlines and the corresponding $u_{rms}$ distribution at (*BR*, *β*, $L^2$, *Wi*) = (75%, 0.59, 40000, 9) are presented in figure A5(a) and A5(b), respectively. The simulation is performed based on the linear extrapolation boundary condition for the tensor **b** on the cylinder wall. The flow asymmetry could be clearly seen in the $u_{rms}$ distribution in figure A5. The flow asymmetry parameter $I$ is also computed and compared. $I$ is 0.0423 for the liner extrapolation boundary condition and 0.0335 for the no flux boundary condition. Compared to that for the no-flux boundary condition, flow fluctuation is more severe for the liner extrapolation boundary, which results in more additional elongational viscosity and thus larger drag force. The comparison indicates that the results obtained by the no flux boundary condition on the cylinder wall are closer to the experimental observations, which is therefore adopted by the present study.

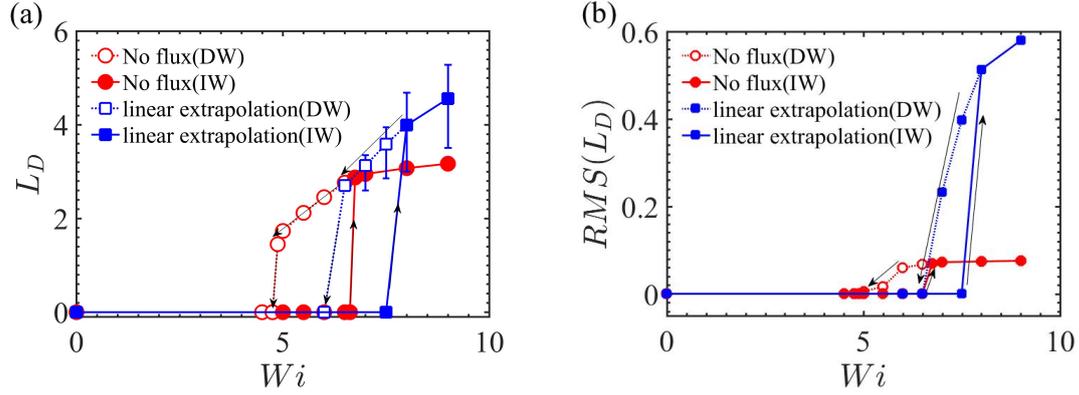

Figure A4. Variation of (a) $L_D$ and (b) the root mean square $L_D$ with $Wi$ at $(BR, \beta, L^2)$ = (75%, 0.59, 40000) for different boundary conditions of the tensor **b** on the cylinder wall. IW and DW denote the increasing and decreasing $Wi$ processes, respectively.

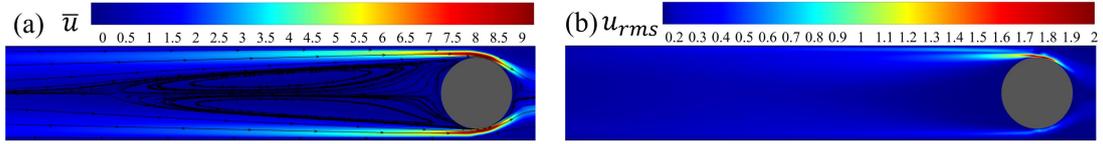

Figure A5. (a) The time averaged $u$-velocity contours and streamlines, and (b) the corresponding $u_{rms}$ distribution at $(BR, \beta, L^2, Wi)$ = (75%, 0.59, 40000, 9). The linear extrapolation boundary condition for the tensor **b** is imposed on the cylinder wall.